\documentclass[conference]{IEEEtran}

\usepackage[noadjust]{cite}

\usepackage{subfigure}
\usepackage{amsmath}
\usepackage{amsthm}
\usepackage{graphicx} % Required for including pictures
\usepackage{caption}
\usepackage{graphicx,subfigure}
\usepackage{multicol} % Required for splitting text into multiple columns
\usepackage{graphicx} % Required for including pictures
\usepackage{amssymb}
\usepackage{float} % Allows putting an [H] in \begin{figure} to specify the exact location of the figure
\usepackage{wrapfig} % Allows in-line images such as the example fish picture
\usepackage{booktabs}
\usepackage{lipsum} % Used for inserting dummy 'Lorem ipsum' text into the template
\usepackage{indentfirst}
\usepackage{bm}
\usepackage{color}
\usepackage{algorithm}
\usepackage{algorithmic}
\usepackage{amsmath}
\usepackage{enumerate}
\usepackage{extarrows}
\usepackage{url}

\usepackage{graphicx}
\usepackage{epstopdf}
\usepackage{multirow}
\usepackage{tablefootnote}
\usepackage[flushleft]{threeparttable}
\usepackage{slashbox}
\usepackage{array}
\usepackage{amsmath}
\usepackage{enumitem}

\IEEEcompsocitemizethanks
\IEEEcompsocthanksitem
\IEEEoverridecommandlockouts

\newtheorem{problem}{\textbf{Problem}}

\newtheorem{corollary}{Corollary}

\newtheorem{theorem}{\textbf{Theorem}}

\newtheorem{question}{Question}

\def\plan{\mathcal{T}}

\def\dcap{Q}
\def\pcap{\Pi}
\def\adfee{\pi}

\def\discount{\delta}

\begin{document}
	
\IEEEoverridecommandlockouts

\title{Economic Viability of Data Trading with Rollover}

\author{
\IEEEauthorblockN{
	Zhiyuan Wang\IEEEauthorrefmark{1},
	Lin Gao\IEEEauthorrefmark{3},
	Jianwei Huang\IEEEauthorrefmark{1}\IEEEauthorrefmark{2}, and
	Biying Shou\IEEEauthorrefmark{4}
	\thanks{This work is supported by the General Research Fund CUHK 14219016 established under the University Grant Committee of the Hong Kong Special Administrative Region, China. This work is also supported by the General Research Fund (CityU 11527316) and HK RGC Theme-based Research Scheme No. T32-101/15-R. This work is also supported by the National Natural Science Foundation of China under Grant 61771162. }
}
		
\IEEEauthorblockA{\IEEEauthorrefmark{1}Department of Information Engineering, The Chinese University of Hong Kong, Hong Kong}
\IEEEauthorblockA{\IEEEauthorrefmark{2}School of Science and Engineering, The Chinese University of Hong Kong, Shenzhen, China}
\IEEEauthorblockA{\IEEEauthorrefmark{3}School of Electronic and Information Engineering, Harbin Institute of Technology, Shenzhen, China}
\IEEEauthorblockA{\IEEEauthorrefmark{4}Department of Management Sciences, City University of Hong Kong, Hong Kong}
}
	
\maketitle
	
\begin{abstract}
Mobile Network Operators (MNOs) are providing more  flexible wireless data services to attract subscribers and increase revenues.
For example, the \textit{data trading market} enables user-flexibility by allowing users to sell leftover data to or buy extra data from each other.
The \textit{rollover mechanism} enables  time-flexibility by allowing a user to utilize his own leftover data from the previous month in the current month.
In this paper, we investigate the economic viability of offering  the data trading market together with the  rollover mechanism, to gain a deeper understanding of the interrelationship  between the user-flexibility and the time-flexibility.
We formulate the interactions between the MNO and mobile users as a multi-slot dynamic game.
Specifically, in each time slot (e.g., every day), the MNO first determines the selling and buying prices with the goal of revenue maximization, then each user decides his trading action (by solving a dynamic programming problem) to maximize his long-term payoff.
Due to the availability of monthly data rollover, a user's daily trading decision corresponds to a dynamic programming problem with two time scales (i.e., day-to-day and month-to-month).
Our analysis reveals an optimal trading policy with a target interval structure, specified by a buy-up-to threshold and a sell-down-to threshold in each time slot.
Moreover, we show that the rollover mechanism makes users sell less and buy more data given the same trading prices, hence it increases the total demand while decreasing the total supply in the data trading market.
Finally, numerical results based on real-world data unveil that the time-flexible rollover mechanism plays a positive role in the user-flexible data trading market, increasing the MNO's revenue by 25\% and all users' payoff by 17\% on average.
\end{abstract}

\IEEEpeerreviewmaketitle

\section{Introduction}
\subsection{Background and Motivation} 
Mobile Network Operators (MNOs) typically offer various wireless data plans. 
The most widely used data plan is a three-part tariff defined by a data cap, a monthly one-time subscription fee, and a linear price for any usage exceeding the data cap \cite{sen2013survey}.
Due to the stochastic nature of users' data demand, the fixed monthly data cap occasionally ends up with a \textit{data leftover} (waste)  in low-demand months or an \textit{overage usage} (additional fee) in high-demand months, both of which users hope to avoid. 
To attract more subscribers  (and further increase revenue), many  MNOs have been exploring various innovations on the three-part tariff data plans to offer more flexibility and reduce users' data consumption uncertainty over time.
Two successful examples are the \textit{rollover mechanism} and the \textit{data trading market}.

The rollover mechanism offers the \textit{time-flexibility} by allowing a user's unused data from the previous month to be used in the current month.
Many MNOs (e.g., AT\&T \cite{ATTrollover}, China Unicom Hong Kong \cite{CUHKrollover}, and China Mobile \cite{CMrollover}) have adopted the rollover data plan since 2015. 
%More specifically, AT\&T requires that the rollover data should be consumed after the monthly data cap, while both China Uncom Hong Kong and China Mobile allow the rollover data consumed prior to the month data cap.
Our previous study in \cite{Zhiyuan2018TMC} has shown that the time-flexible rollover mechanism leads to a win-win situation.
That's, it enables users to hold the leftover data in low-demand months to compensate the heavy demand in the future.
It also allows the MNO to extract more consumer surplus through a higher subscription fee.
As a result, the rollover mechanism increases the total social welfare.

Different from the rollover mechanism, the data trading market promotes the \textit{user-flexibility}.
In 2014, China Mobile Hong Kong (CMHK) launched the first data trading market, called the 2nd exChange Market (2CM) \cite{CMHK2cm}.
It allows subscribers to sell leftover data to or buy extra data from others.
The transaction prices may change over time, depending on the supply and demand relationship in the data trading market.
In addition, CMHK acts as the middleman and benefits from the difference between the buying price and selling price.
It is shown in \cite{zheng2015secondary} that the trading market is beneficial to the MNO, since its revenue gain from the trading market is larger than the revenue loss from overage fee. 
%Despite the booming data trading market, CMHK has not yet implemented the widely-used rollover mechanism.

Although the two mechanisms are increasingly popular, the economic viability of the data trading market with a rollover mechanism has not been clearly demonstrated, which may impede the MNO's joint adoption in the telecom market.
\begin{figure}
	\centering
	\setlength{\abovecaptionskip}{5pt}
	\setlength{\belowcaptionskip}{0pt}
	\includegraphics[width=0.85\linewidth]{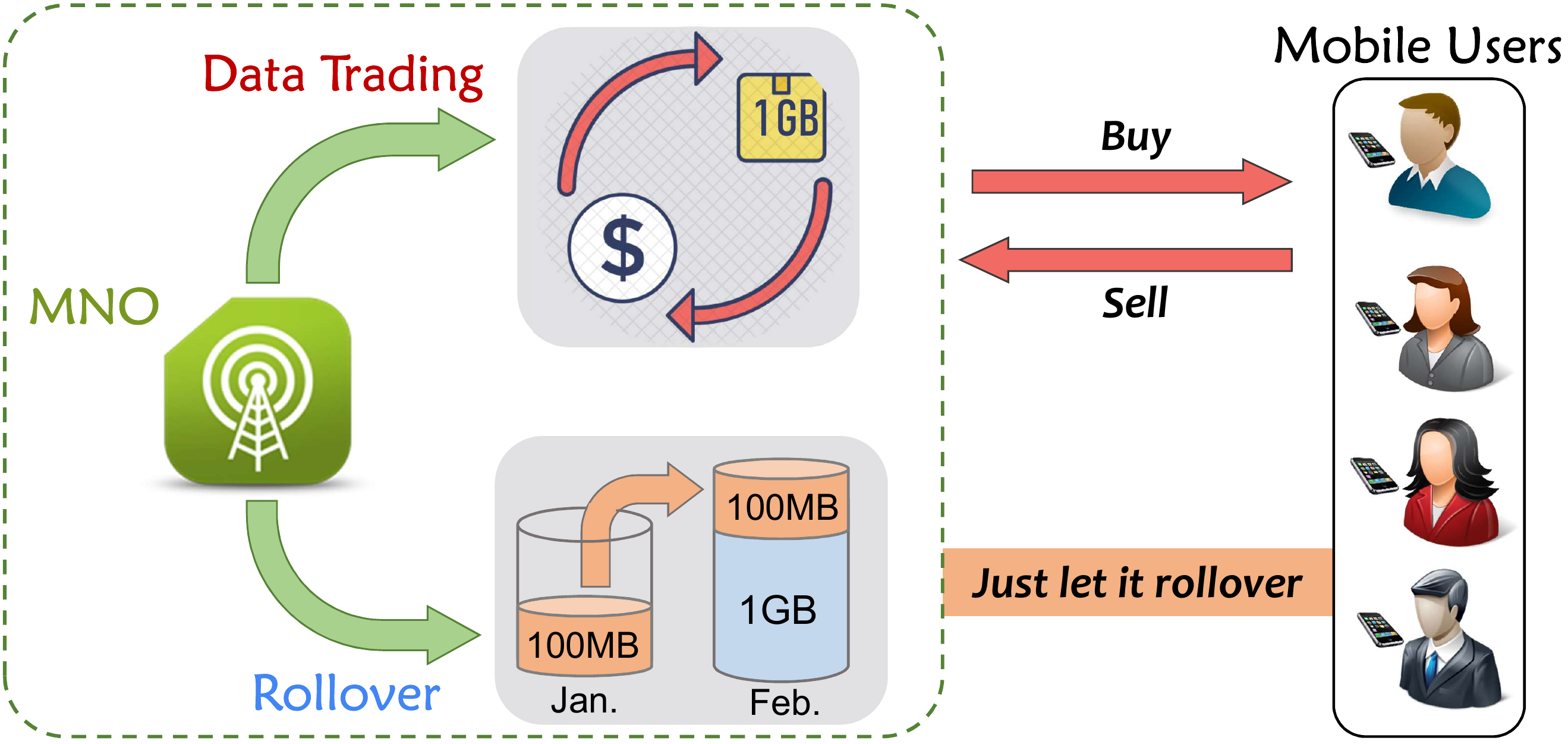}
	\caption{The data trading market with the rollover mechanism.\vspace{-8pt}}
	\label{fig:Application}
\end{figure}

First, the rollover mechanism offers users a conservative way of managing their leftover data, i.e., holding for future use, while the data trading encourages a more radical attitude towards data manipulation by taking advantage of the time-variant trading prices.
It is easy to imagine that the rollover mechanism will remarkably change users' data trading behaviors, which motivates us to ask the first question:
\begin{question}
	How will users adjust their selling and buying decisions under the rollover mechanism?
\end{question}

Second, although the joint adoption of the rollover mechanism and the data trading market offers more flexibility to  mobile users, it is not yet clear whether the joint adoption is also beneficial for the MNO, comparing with adopting them separately.
This motivates us to ask the second question:
\begin{question}
	Will rollover mechanism increase or reduce the MNO's benefit from data trading market?
\end{question}

%To address Question 2, it is critical to understand the fundamental relation between the time-flexibility and user-flexibility. 

In this paper, we will study the economic viability of the data trading market with a rollover mechanism (as in Fig. \ref{fig:Application}) and address the above two key questions.
As far as we know, the interrelationship  between the time-flexibility and the user-flexibility has not been studied before.
We hope that this paper could lead to a much better understanding about them.

\subsection{Key Results and Contributions} 
%By studying the interactions between the MNO and mobile users, we reveal the impact of the rollover mechanism on the data trading market.
Our main results and key contributions are as follows:
\begin{itemize} 
	%====================================================================================
	\item \textit{Viability Analysis on the Data Trading Market with a Rollover Mechanism:} 
	We formulate the interactions between the MNO and users as a multi-slot dynamic game.
	In each time slot (e.g., every day), the MNO first determines the selling price and buying price with the goal of revenue maximization, then each user determines his trading action to maximize his long-term payoff.
	Due to the availability of monthly data rollover, a user's optimal trading decision corresponds to a dynamic programming problem with two time scales (e.g., day-to-day and month-to-month).
	Despite the complexity of the model, we are able to explicitly characterize the optimal decisions for the MNO and users.
	%====================================================================================
	\item \textit{Optimal Trading Decision:}
	We analyze users' trading problem through backward induction and unveil a target interval trading policy specified by a buy-up-to threshold and a sell-down-to threshold in each time slot.
	Such a target interval policy allows simple implementation for users in practice and it recovers the optimal trading decision without the rollover mechanism as a special case.
	%====================================================================================
	\item \textit{Impact of Rollover Mechanism:}
	By comparing the thresholds of the target interval policy, we find that the rollover mechanism makes users sell less and buy more data under the same trading prices, which allows users to hold more data to avoid the potential overage fee and wait for higher selling prices.
	Hence the rollover mechanism increases the total trading market demand and reduces the total supply.	
	\item \textit{Performance Evaluation based on Real-world Data:} 
	We use a group of mobile users' six-month daily data usage traces to  evaluate the performance improvement brought by the rollover mechanism.
	The numerical results show that offering rollover service can increase the MNO's average revenue by 25\% and users' average payoff by 17\%, which demonstrate that the rollover mechanism can be very beneficial to the data trading market.
\end{itemize}

\subsection{Related Literature} 
This work focuses on the interrelationship of two data pricing mechanisms, i.e., the \textit{data trading market} and the \textit{rollover mechanism}.
For clarity we list in Table \ref{table: Related Literatures} the related literature in terms of the mechanism and main methodologies.
%We also point out how our work differs from these studies.

There are many excellent works on the traditional mobile data market without considering the two new mechanisms.
Some focused on the optimization of pricing (e.g., \cite{ha2012tube,ma2016time}) and data cap (e.g., \cite{ma2016usage}) from the MNO's perspective.
Other studies (e.g., \cite{andrews2014understanding,Zheng2018dynamics}) examined a single user's data consumption dynamics using multi-period models.
%However, all of these did not consider the newly introduced innovative applications.

Data trading market has been studied in \cite{zheng2015customized,yu2017mobile,andrews2016understanding}. 
Zheng \textit{et al.} in \cite{zheng2015customized} considered the auction-based trading platform.
Yu \textit{et al.} in \cite{yu2017mobile} analyzed the users' behavior with prospect theory.
%Wang \textit{et al.} in \cite{wang2016user} considered the data trading through hot-spot.
Andrews \textit{et al.} in \cite{andrews2016understanding} developed a multi-period model to study a single user's trading actions.
However, these papers did not take into account the rollover mechanism.

The rollover mechanisms have been studied in \cite{zheng2016understanding,wei2018novel,Zhiyuan2017rollover,Zhiyuan2018duopoly,Zhiyuan2018MobiHoc}.
Zheng \textit{et al.} in \cite{zheng2016understanding} found that moderately price-sensitive users can benefit from subscribing to the rollover data plan.
Wei \textit{et al.} in \cite{wei2018novel} focused on the impact of expiration time. 
In our previous works, we studied the consumption priority of the rollover data \cite{Zhiyuan2017rollover}, the MNOs' market competition \cite{Zhiyuan2018duopoly}, and the multi-cap design problem \cite{Zhiyuan2018MobiHoc}.
However, none of above studies considered users' optimal data trading decisions in the data trading market.

Our paper differs from the aforementioned works in the following key aspects:
First, our model generalizes the data trading market and the rollover mechanism into a unified framework with two-dimensional flexibilities (i.e., user and time).
%The additional model complexity is far more than the simple composition of the separate cases.
Second, we explicitly characterize a user's optimal data trading policy in each time slot considering the availability of data rollover.
Third, we examine the viability of the data trading market and the rollover mechanism, which has never been studied before.

The remainder of this paper is organized as follows.
Section \ref{Section: System Model} introduces the system model. 
In Section \ref{Section: Trading Policy}, we study users' trading policy.
In Section \ref{Section: MNO's Pricing}, we analyze the MNO's pricing problem.
Section \ref{Section: numerical results} presents the numerical results  and Section \ref{Section: conclusion} concludes this paper.

\begin{table}
	\setlength{\abovecaptionskip}{1pt}
	\setlength{\belowcaptionskip}{0pt}
	\renewcommand{\arraystretch}{1.04}		
	\caption{Mobile data pricing literatures.} 
	\label{table: Related Literatures}
	\centering
	\begin{tabular}{|c|cc|cc|}
		\hline
		\multirow{ 2}{*}{Reference} 	& \multicolumn{2}{c|}{Application}	& \multicolumn{2}{c|}{Methodology}	\\
		& Rollover	& Data Trading	& Multi-period & Multi-user	 \\
		\hline\hline
		
		%% all no
		\cite{ha2012tube,ma2016time,ma2016usage}
		& $\times$	& $\times$	& $\times$	& $\times$ \\
		% all no + multi-period + data
		\cite{andrews2014understanding,Zheng2018dynamics}	
		& $\times$ 	& $\times$	& $\surd$	& $\times$ \\
		
		\hline
		
		%% trading + muiti-user
		\cite{zheng2015customized,yu2017mobile} 				
		& $\times$	& $\surd$	& $\times$	& $\surd$ \\
		%% trading + multi-period
		\cite{andrews2016understanding}	
		& $\times$	& $\surd$	& $\surd$	& $\times$ \\
		
		\hline
		
		% rollover
		\cite{zheng2016understanding}		
		& $\surd$	& $\times$	& $\times$	& $\times$ \\
		% rollover + data
		\cite{wei2018novel,Zhiyuan2017rollover,Zhiyuan2018duopoly,Zhiyuan2018MobiHoc}	
		& $\surd$	& $\times$	& $\times$	& $\surd$ \\
		
		\hline
		
		% this paper
		\textbf{This Paper}					
		& $\bm\surd$ 	& $\bm\surd$	& $\bm\surd$	& $\bm\surd$		\\
		\hline
	\end{tabular}	\vspace{-5pt}
\end{table}

\section{System Model}\label{Section: System Model}
We consider a telecom market where a set $\mathcal{N}=\{1,2,..,N\}$ of mobile users subscribe to a Mobile Network Operator (MNO).
The MNO offers a three-part tariff data plan together with the \textit{rollover} and \textit{data trading} services.
The time is slotted with the index $t\in\{1, 2, 3, ...\}$. 
In each slot $t$ (e.g., every day), the MNO determines the trading prices and each user $n\in\mathcal{N}$ takes a trading action (how much to buy or sell). 
The data rollover happens at the end of each billing cycle (i.e., every month), as a result of the user's data consumption and data trading decisions in the month.

Next we first introduce the wireless data services in Section \ref{Subsection: Wireless Data Services}.
Then we formulate users' trading problem and MNO's pricing problem in Section \ref{Subsection: Mobile User Modeling} and Section \ref{Subsection: MNO Pricing}, respectively.

\subsection{Wireless Data Services}\label{Subsection: Wireless Data Services}
We introduce the MNO's wireless data services from the following four aspects.
%: the mobile data plan, the rollover mechanism, and the data trading market.

\subsubsection{Mobile Data Plan}
We characterize a mobile data plan by a tuple $\plan=\{\dcap,\pcap,\adfee\}$.
The user pays a \textit{monthly subscription fee} $\pcap$ for the data consumption up to the \textit{data cap} $\dcap$.  
For unit data consumption exceeding the data cap, the user pays the \textit{overage fee} $\adfee$.
Such a tuple $\plan$ includes both  the pure usage-based data plan (i.e., $\dcap=0$ and $\pcap=0$) and the unlimited data plan (i.e., $\dcap=+\infty$) as special cases.

In practice, mobile users usually sign a one-year or two-year contract with the MNO for a particular data plan.
%\footnote{In practice, users' billing cycles may not be synchronous. For example, MNOs in Hong Kong allow users to start the contract in any day of the calendar month. However, the MNOs in mainland China require the billing cycle be synchronous with the calendar month. For notation clarity, we will focus on the synchronous case in the main paper and discuss the differences in our technical report \cite{Technicalreport}.}
We consider a time horizon consisting of $M$ months and suppose that there are a total of $K$ time slots in each month (e.g., 30 days).
Denote $m\in\{1,2,...,M\}$ as the $m$-th month and $k\in\{1,2,...,K\}$ as the $k$-th time slot in a particular month.
Hence the time slot $(m,k)$ corresponds to $t=K(m-1)+k$.
For the sake of presentation, we use $(m,k)$ and $t$ interchangeably.

%We characterize the contract period in two time scales. 
%More specifically, we consider a user signing a contract of $M$ months (e.g., 24 months) with the MNO.
%Without loss of generality, we suppose that there are a total of $K$ time slots in each month (e.g., 30 days or 720 hours).\footnote{In practice, the number of time slots in different billing cycles are distinct, e.g., 31 days in January and 28 days in February. We assume the same number of time slots for simplicity, since it does not change our key results.}
%
%
%As shown in Fig. \ref{fig:TimeScale}, we index the time scale in a \textbf{\textit{descending order}} and denote $m\in\{1,2,...,M\}$ as the $m$-th month to the contract end and $k\in\{1,2,...,K\}$ as the $k$-th slot to the end of the current month.
%In this case, $(M,K)$ represents the start of the contract period, while $(1,1)$ is the end.
%Accordingly, at time slot $(m,k)$, there are $t=K(m-1)+k$ time slots to the contract end.
%In the following, we will use $(m,k)$ and $t$ interchangeably.

\subsubsection{Rollover Mechanism}
The rollover mechanism allows a user's leftover data from the previous month to be used in the current month.
Different  rollover mechanisms can be classified based on the consumption priority between the rollover data and the current monthly data cap, the impact of which has been extensively studied in \cite{Zhiyuan2017rollover}.
In this work, we focus on one of the most common implementations by MNOs (including China Mobile and China Unicom HK): \textit{the rollover data from the previous month is consumed prior to the current monthly data cap and expires at the end of the current month}.

\subsubsection{Data Trading Market}
Mobile users can sell their leftover data or buy extra data in the data trading market.
The MNO determines the selling price $p_{t}^{s}$ and the buying price $p_{t}^{b}$ based on the total trading market supply (from sellers) and the total trading market demand (from buyers) in each time slot $t$.
In this work, we consider the MNO's revenue-maximizing pricing strategy, i.e., the MNO decides the trading prices $\bm{p}_{t}=\{p_t^s,p_{t}^{b}\}$ to maximize its revenue in slot $t$.
Our later analysis in Section \ref{Section: MNO's Pricing} shows that the revenue-maximizing pricing clears the data trading market, i.e., the trading market demand equals to the trading market supply.
Other objectives (e.g., maximizing the long-term revenue) are also possible but may not clear the market efficiently, which will be discussed in Section \ref{Section: conclusion}.

\begin{figure}
	\centering
	\setlength{\abovecaptionskip}{3pt}
	\setlength{\belowcaptionskip}{0pt}
	\includegraphics[width=0.95\linewidth]{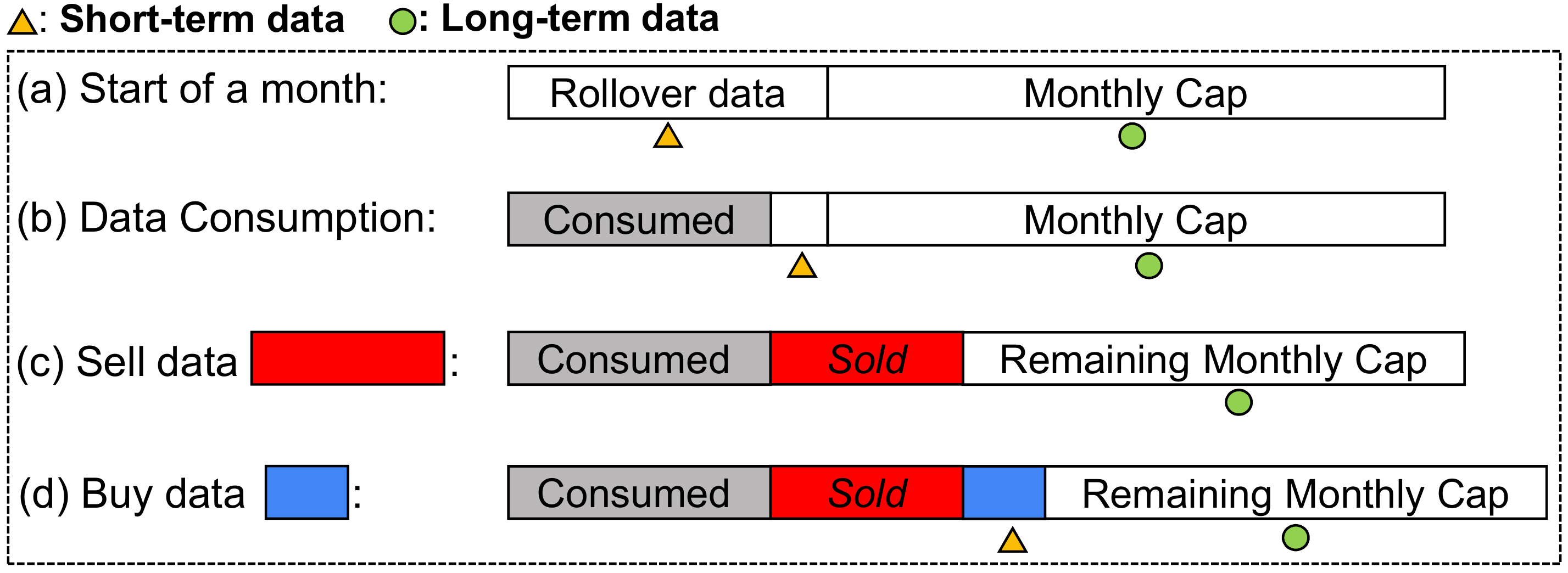}
	\caption{The MNO's policy on rollover and data trading.}
	\label{fig: Priority_Roll_Trade}
\end{figure}
\subsubsection{Implementation of Data Rollover \& Trading}\label{Subsubsection: Rollover Trading}
There will be some new considerations for the MNO when offering  both rollover and trading services.
Under the rollover mechanism, users have two types of data caps that differ in  terms of the expiration time, which we call the \textit{short-term} data cap and the \textit{long-term} data cap.
\begin{itemize}
	\item The \textit{shot-term data cap} can be consumed in the current month, and expires at the end of the current month.
			
	\item The \textit{long-term data cap} not only can be consumed in the current month, but also can rollover to the next month.
			
\end{itemize}

Data trading behavior affects the short-term and long-term data caps.
When a user buys some extra data from other users, most MNOs (e.g., AT\&T and China Mobile) require that \textit{the purchased data cannot rollover} (i.e., it is short-term) and \textit{is consumed with the top priority}.
When a user wants to sell some extra data to other users, we assume that \textit{the short-term data is sold prior to the long-terms data}, which will lead to the maximum flexibility to the user.

To sum up, the short-term data includes two parts: the rollover data from the previous month and the purchased data from the trading market in the current month.
The long-term data only corresponds to the current monthly data cap $\dcap$.
Moreover, the short-term data is consumed and sold prior to the long-term one.
Fig. \ref{fig: Priority_Roll_Trade} provides an illustration of the MNO's policy of data rollover and trading.

\subsection{Mobile Users' Decisions}\label{Subsection: Mobile User Modeling}
We introduce how to model users' decisions in five aspects.
%We first introduce the data trading action and the data demand realization, then we derive the user's one-slot payoff.

\subsubsection{Data Volume}
We use $\{\dcap^{n},\pcap^{n},\adfee\}$ to represent the data plan of user $n$.
Note that the per-unit fee $\adfee$ is usually the same among various  data plans from the same MNO \cite{sen2013survey}.
In time slot $t$, we denote $e_{t}^{n}$ and $q_{t}^{n}$ as the short-term data volume and the long-term data volume, respectively. 
Since $\dcap^{n}$ is the potential maximal long-term data, $q_{t}^{n}\le\dcap^{n}$.
We further denote $\bm{e}_t=\{e_t^n, n\in\mathcal{N}\}$ and $\bm{q}_t=\{q_t^n, n\in\mathcal{N}\}$ as all users' short-term data vector and long-term data vector, respectively.

\begin{figure}
	\centering
	\setlength{\abovecaptionskip}{3pt}
	\setlength{\belowcaptionskip}{0pt}
	\includegraphics[width=0.95\linewidth]{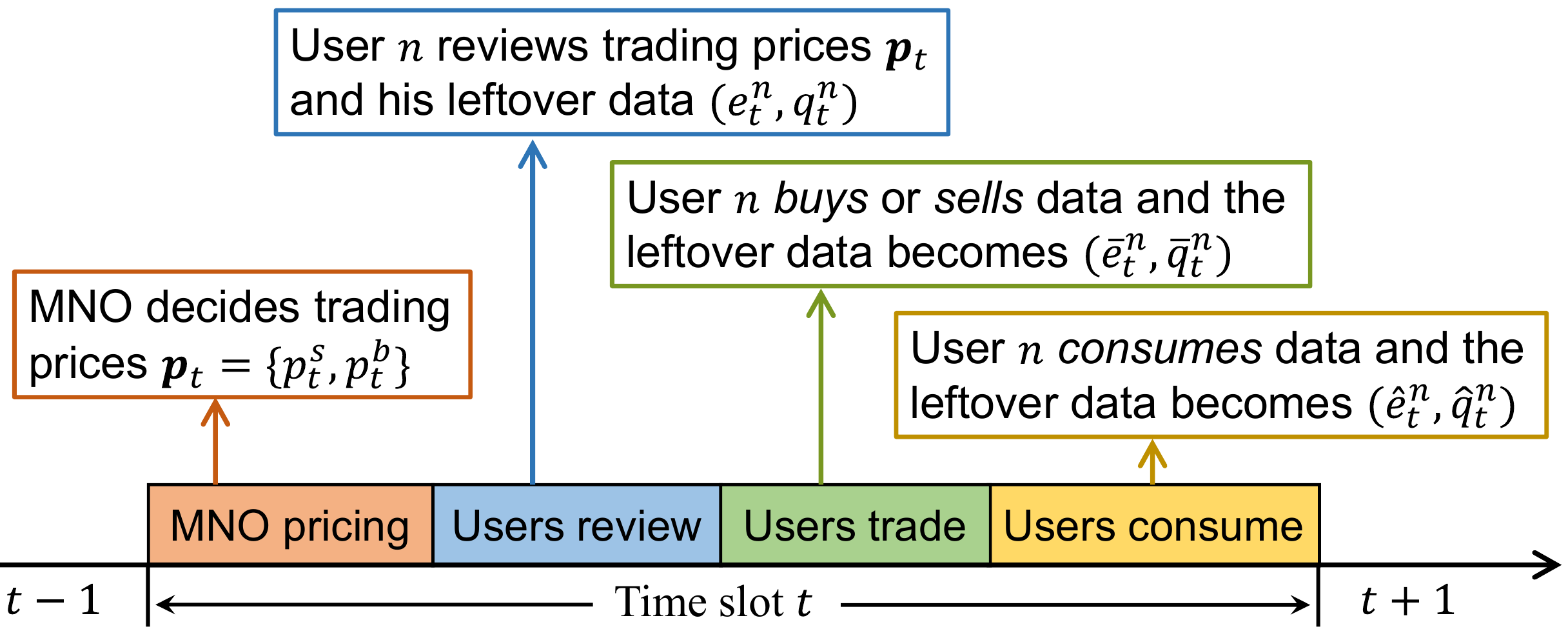}
	\caption{The sequence of events in each time slot $t$. }
	\label{fig: OneSlot}
\end{figure}

The sequence of events in each time slot $t$ is  as follows (also see Fig. \ref{fig: OneSlot}):
\begin{itemize}
	\item \textbf{MNO Pricing:} The MNO decides the prices $\bm{p}_{t}=\{p_{t}^{s},p_{t}^{b}\}$.
	\item \textbf{Users Review:} Each user $n\in\mathcal{N}$ reviews his leftover data volume $(e_{t}^{n},q_{t}^{n})$ and the trading prices $\bm{p}_{t}=\{p_{t}^{s},p_{t}^{b}\}$.
	\item \textbf{Users Trade:} Each user $n$ takes a trading action (i.e., buy, sell, or no trading) and his leftover data becomes $(\bar{e}_{t}^{n},\bar{q}_{t}^{n})$.
	\item \textbf{Users Consume:} After user $n$ consumes data, his leftover data volume decreases to $(\hat{e}_{t}^{n},\hat{q}_{t}^{n})$ at the end of slot $t$.
\end{itemize}

Next we explain users' trading actions and data consumptions.
We adopt the notations of $(\cdot)^+=\max\{\cdot,0\}$ and $(\cdot)^-=\min\{\cdot,0\}$ for brevity.

\subsubsection{Trading Action}
We denote $a_{t}^{n}$ as the data trading action of user $n$ in time slot $t$.
It covers the following three cases.
\begin{itemize} 
	\item Buy data ($a_{t}^{n}>0$): User $n$ buys $a_{t}^{n}$ amount of data with a  unit-price $p_{t}^{b}$.
			His short-term data volume increases, i.e., $\bar{e}_{t}^{n}=e_{t}^{n}+a_{t}^{n}$, while the long-term data volume does not change, i.e., $\bar{q}_{t}^{n}=q_{t}^{n}$.
	\item Sell data ($a_{t}^{n}<0$): User $n$ sells $|a_{t}^{n}|$ amount of data with a  unit-price $p_{t}^{s}$.
			His short-term data is sold first, i.e., $\bar{e}_{t}^{n}=(e_{t}^{n}+a_{t}^{n})^+$. 
	If that is not enough to, then he will sell the long-term data, i.e., $\bar{q}_{t}^{n}=q_{t}^{n}+(e_{t}^{n}+a_{t}^{n})^-$. 
%	\footnote{Here $(x)^-=\min\{0,x\}$.}
	\item No trading ($a_{t}^{n}=0$): User $n$ does not sell or buy data.
			His data volume remains the same, i.e., $\bar{e}_{t}^{n}=e_{t}$ and $\bar{q}_{t}^{n}=q_{t}^{n}$.
\end{itemize}

To sum it up, user $n$'s  leftover short-term and long-term data volumes after data trading can be represented as:
\begin{equation}\label{Equ: (e,q) after trade}
	\left\{
	\begin{aligned}
		&\bar{e}_{t}^{n} = (e_{t}^{n}+a_{t}^{n})^+,	\\
		&\bar{q}_{t}^{n} = q_{t}^{n}+(e_{t}^{n}+a_{t}^{n})^-, 
	\end{aligned}
	\right.		
\end{equation}
where $a_{t}^{n}\ge-(e_{t}^{n}+q_{t}^{n})$, since user $n$ can sell at most $e_{t}^{n}+q_{t}^{n}$ units of data.
For notation simplicity, we denote $\bar{z}_{t}^{n}$ as the total data volume of user $n$ after his trading action $a_{t}^{n}$, given by
\begin{equation}\label{Equ: z after trade}
\begin{aligned}
\bar{z}_{t}^{n} = \bar{e}_{t}^{n}+\bar{q}^n_t = e_{t}^{n}+q^n_t + a_{t}^{n}.
\end{aligned}
\end{equation}

%{\color{red}
%Furthermore, when user $n$ takes the trading action $a_{t}^{n}$, his total volume updates as follows: 
%\begin{equation}\label{Equ: z after trade a}
%\bar{z}_{t}^{n} = z_{t}^{n} + a_{t}^{n} ,
%\end{equation}
%which implies a mapping from $a_{t}^{n}$ to  $\bar{z}_{t}^{n}$.
%For notation convenience, we sometimes use $\bar{z}_{t}^{n}$ to represent the trading action of user $n$ instead of $a_{t}^{n}$.
%Accordingly, we substitute (\ref{Equ: z after trade a}) into (\ref{Equ: (e,q) after trade}) and obtain
%\begin{equation}\label{Equ: (e,q) after trade z}
%\left\{
%\begin{aligned}
%&\bar{e}_{t}^{n} = \left(\bar{z}_{t}^{n}-q_{t}^{n}\right)^+,		\\
%&\bar{q}_{t}^{n} = q_{t}^{n}+\left(\bar{z}_{t}^{n}-q_{t}^{n}\right)^- .
%\end{aligned}
%\right.		 
%\end{equation}
%
%We illustrate how the short-term data volume and long-term data volume change according to $\bar{z}_{t}^{n}$ in Fig. \ref{fig: Transition_Trade}.
%}

\subsubsection{Data Consumption}
As each user $n$'s data consumption in time slot $t$ is stochastically random (i.e., not known by the user beforehand), we model it  as a random variable $x_{t}^{n}$ with a PDF $f_n(\cdot)$.
The distribution information is available from the historical data consumption record. 
In Section \ref{Section: numerical results}, we will use real-world data to estimate the distribution $f_n(\cdot)$ for each user.

%Denote $(\hat{e}_{t}^{n},\hat{q}_{t}^{n})$ as the leftover data volume of user $n$ after he consumes $x_{t}^{n}$ units of data.
According to the MNO's policy in Section \ref{Subsubsection: Rollover Trading}, user $n$ consumes his short-term data  first, i.e., $\hat{e}_{t}^{n}=( \bar{e}_{t}^{n}-x_{t}^{n} )^+ $. 
If the short-term data is not enough, then he further consumes the long-term data. 
After the long-term data decreases to zero, further data consumption  will lead to an overage charge with a unit price $\adfee$.
Therefore, the leftover short-term and long-term data volumes after data consumption are
\begin{equation}\label{Equ: (e,q) after demand}
	\left\{
	\begin{aligned}
		& \hat{e}_{t}^{n}=( \bar{e}_{t}^{n}-x_{t}^{n} )^+ ,			\\
		& \hat{q}_{t}^{n}=\left( \bar{q}_{t}^{n} + (\bar{e}_{t}^{n}-x_{t}^{n})^- \right)^+.
	\end{aligned}
	\right.
\end{equation}

%{\color{red}
%Similarly, we substitute (\ref{Equ: (e,q) after trade z}) into (\ref{Equ: (e,q) after demand}) and express $(\hat{e}_{t}^{n},\hat{q}_{t}^{n})$ by $\bar{z}_{t}^{n}$ and $x_t^n$, as follows:
%\begin{equation}\label{Equ: after demand z}
%	\left\{
%	\begin{aligned}
%		& \hat{e}_{t}^{n} = \left( \bar{z}_{t}^{n}-q_{t}^{n} -x_{t}^{n} \right)^+ ,								\\
%		& \hat{q}_{t}^{n} = \left(\bar{z}_{t}^{n}-x_{t}^{n}\right)\big|_{0}^{q_{t}^{n}},	
%	\end{aligned}
%	\right.
%\end{equation}
%which is illustrated in Fig. \ref{fig: Transition_TradeDemand}.
%}

%\begin{figure} 
%	\centering 
%	\setlength{\abovecaptionskip}{0pt}
%	\setlength{\belowcaptionskip}{0pt}
%	\subfigure[$(\bar{e}^n_t,\bar{q}^n_t)$ vs $\bar{z}^n_t$]{\label{fig: Transition_Trade}\includegraphics[width=0.45\linewidth]{Transition_Trade.png}} \quad
%	\subfigure[$(\hat{e}^n_t,\hat{q}^n_t)$ vs $\bar{z}^n_t$]{\label{fig: Transition_TradeDemand}\includegraphics[width=0.45\linewidth]{Transition_TradeDemand.png}} 
%	\caption{Illustration of $(\bar{e}_{t}^{n},\bar{q}_{t}^{n})$ and $(\hat{e}_{t}^{n},\hat{q}_{t}^{n})$.}
%\end{figure}

\subsubsection{One-slot User Payoff}
%Before deriving users' payoffs, we summarize the sequence of events in each time slot $t$ as follows.
%At the beginning of the slot $t$, the trading prices $\bm{p}_{t}=\{p_{t}^{s},p_{t}^{b}\}$ are realized and the leftover data volume $(e_{t}^{n},q_{t}^{n})$ is reviewed by user $n$.
%The user $n$ then makes his data trading decision $\bar{z}_{t}^{n}$ and his leftover data volume is adjusted to $(\bar{e}_{t}^{n},\bar{q}_{t}^{n})$ based on (\ref{Equ: (e,q) after trade z}).
%Finally, his random demand $x_{t}^{n}$ is realized and the leftover data volume is reduced to $(\hat{e}_{t}^{n},\hat{q}_{t}^{n})$ based on (\ref{Equ: after demand z}).

A user's one-slot payoff  depends on the one-slot utility from consuming data, overage charge, and the trading income or cost.
First, we use a general utility function $u_n(x)$ to represent user $n$'s one-slot utility of consuming $x$ units data. 
Function $u_n(x)$ is assumed to be an increasing and concave function in $x$.
Second, user $n$ has to pay the overage charge $\adfee(x^n_t-e^n_t-q^n_t-a^n_t)^+$ if his data demand exceeds the leftover total data volume after trading.
Third, the data trading decision $a_{t}^{n}$ brings a monetary income $p_{t}^{s}\cdot(-a^n_t)^+$ or a monetary cost $p_{t}^{b}\cdot(a^n_t)^+$.
Therefore, the \textit{one-slot payoff} of user $n$ is 
\begin{equation}
\begin{aligned}
v_{t}^{n}\left( e_t^n,q_t^n,\bm{p}_t,a^n_t, x_t^n \right) =& u_n(x_{t}^{n}) -\adfee\left( x_{t}^{n}- e^n_t-q^n_t-a^n_t \right)^+ \\
& + p_{t}^{s} \cdot (-a^n_t)^+ -p_{t}^{b} \cdot (a^n_t)^+.
\end{aligned}
\end{equation}
%where $x^n_t$ is the user's data consumption.

We take the expectation over the random data consumption $x_{t}^{n}$ to derive user $n$'s  \textit{one-slot expected payoff}:
\begin{equation}
\bar{v}_{t}^{n}\left(  e_t^n,q_t^n,\bm{p}_t,a^n_t \right) 
= \int_{0}^{+\infty} v_{t}^{n}\left( e_t^n,q_t^n,\bm{p}_t,a^n_t, x  \right) f_n(x){\rm d}x .	
\end{equation}

%For notation simplicity, we denote $\bar{z}_{t}^{n}$ as the total data volume of user $n$ after his trading action $a_{t}^{n}$, given by
%\begin{equation}\label{Equ: z after trade}
%\begin{aligned}
%\bar{z}_{t}^{n} = e_{t}^{n}+q^n_t + a_{t}^{n}.
%\end{aligned}
%\end{equation}

To illustrate the key insights of the optimal trading policy (in Section \ref{Section: Trading Policy}), we treat $\bar{z}^n_t$ (defined in (\ref{Equ: z after trade})) as the trading decision variable of user $n$, instead of using $a^n_t$.
%This can provide a understanding on the optimal trading policy which will be analyzed in Section \ref{Section: Trading Policy}.
Hence we rewrite the user's \textit{one-slot expected payoff} as follows:
\begin{equation}\label{Equ: one-slot expected payoff}
\bar{v}_{t}^{n}\left(  e_t^n,q_t^n,\bm{p}_t ,\bar{z}^n_t\right) = W(\bar{z}_{t}^{n}) + J(e_{t}^{n}+q^n_t-\bar{z}_{t}^{n},\bm{p}_{t}),
\end{equation}
where $W_n(\bar{z})$ and $J(z,\bm{p}_{t})$ are given by
\begin{equation}
W(\bar{z}) \triangleq \int_{0}^{+\infty} \left[ u_n(x) -\adfee\left( x-\bar{z} \right)^+ \right]f_n(x){\rm d}x,
\end{equation}
\begin{equation}
J(z,\bm{p}_{t}) \triangleq p_{t}^{s} \cdot(z)^+ - p_{t}^{b} \cdot (-z)^+.
\end{equation}

\subsubsection{Multi-slot Data Trading \& Rollover}
Based on user $n$'s  one-slot expected payoff in (\ref{Equ: one-slot expected payoff}), we will further formulate a user's multi-slot data trading problem.
Before that, we first introduce the transition between consecutive time slots, which contains the following two cases:
\begin{itemize} 
	\item If the current time slot $t$ \textit{is not} the end of a month, then the user's data volume at the beginning of the next time slot equals to that at the end of the current time slot, i.e., 
	\begin{equation}\label{Equ: transition inner}
	\left\{
	\begin{aligned}
	& e_{t+1}^{n}=\hat{e}_{t}^{n},\\
	& q_{t+1}^{n}=\hat{q}_{t}^{n}.
	\end{aligned}
	\right.
	\end{equation}
	\item If the current time slot $t$ \textit{is} the end of a month, then the $\hat{e}_{t}^{n}$ units of short-term data expires, while the $\hat{q}_{t}^{n}$ units of long-term data will rollover and becomes the short-term data of the next month.
	Moreover, the monthly data cap $\dcap^{n}$ becomes available to user $n$ again.
	Therefore, the data volume in the next time slot is
	\begin{equation}\label{Equ: transition across}
	\left\{
	\begin{aligned}
	& e_{t+1}^{n}=\hat{q}_{t}^{n},\\
	& q_{t+1}^{n}=\dcap^{n}.
	\end{aligned}
	\right.
	\end{equation}
	
\end{itemize}

%In the data trading market, user $n$ makes his trading decision based on his current reviewed trading prices $\bm{p}_{t}$ and his prospective  future trading prices.
%We assume that the $\{\bm{p}_{t},\ \forall\ 1\le t\le T\}$ forms a Markov chain, i.e.,
%\begin{equation}
%	\mathbb{P}\left(\bm{p}_{t+1}|\bm{p}_1,\bm{p}_2,...,\bm{p}_{t}\right) =
%	\mathbb{P}\left(\bm{p}_{t+1}|\bm{p}_{t}\right),\ \forall\ 1\le t\le T.
%\end{equation}

We denote $V_{t}^{n}(e_{t}^{n},q_{t}^{n},\bm{p}_{t})$ as user $n$'s maximal expected total discounted payoff from slot $t$ to his contract end, given his current data volume $(e_{t}^{n},q_{t}^{n})$ and the trading prices $\bm{p}_{t}$. 
We also refer to $V_{t}^{n}(e_{t}^{n},q_{t}^{n},\bm{p}_{t})$ as user $n$'s \emph{value function}  in time slot $t$.
Accordingly, we can formulate user $n$'s multi-slot  data trading as the following dynamic programming  problem.
\begin{problem}[User $n$'s Multi-Slot Data Trading Problem] \label{Problem: Optimal Data Trading}
	For user $n$ in the $k$-th time slot of the $m$-th month, i.e., $t=K(m-1)+k$, his value function has  three cases:	
	\begin{enumerate} [leftmargin=13 pt]
		\item[1.] If $m=M$ and $k=K$, the value function is
			\begin{equation}\label{Equ: problem last day of contract}
			V_{t}^{n}(e_{t}^{n},q_{t}^{n},\bm{p}_{t})= \max\limits_{\bar{z}_{t}^{n}\ge0} \big\{J(e_{t}^{n}+q^n_t-\bar{z}_{t}^{n},\bm{p}_{t})+ W(\bar{z}_{t}^{n}) \big\}.
			\end{equation}
		\item[2.] If $m<M$ and $k=K$, the value function is
			\begin{equation}\label{Equ: problem last day of month}
			\begin{aligned}
			V_{t}^{n}(e_{t}^{n},q_{t}^{n},\bm{p}_{t})=
			\max\limits_{\bar{z}_n\ge0} \big\{ & J(e_{t}^{n}+q^n_t-\bar{z}_{t}^{n},\bm{p}_{t})+ W(\bar{z}_t^n) \\
			& +\discount\cdot\mathbb{E}_t \left[V_{t+1}^{n}\left( \hat{q}_{t}^{n},\dcap^{n},\bm{p}_{t+1} \right)\right] \big\}.
			\end{aligned}
			\end{equation}
		\item[3.] If $k<K$, the value function is
		\begin{equation}\label{Equ: problem middle}
		\begin{aligned}
		V_{t}^{n}(e_{t}^{n},q_{t}^{n},\bm{p}_{t})= 
		\max\limits_{\bar{z}_{t}^{n}\ge0} \big\{ & J(e_{t}^{n}+q^n_t-\bar{z}_{t}^{n},\bm{p}_{t})+ W(\bar{z}_{t}^{n}) \\
		&  +\discount\cdot\mathbb{E}_t\left[V_{t+1}^{n}\left(\hat{e}_{t}^{n},\hat{q}_{t}^{n} ,\bm{p}_{t+1} \right) \right]  \big\}.
		\end{aligned}
		\end{equation}
	\end{enumerate}
\end{problem}

Case 1 corresponds to the very last time slot (contract ending  day).
On the RHS, the terms inside the brackets is the one-slot expected payoff.

Case 2 corresponds to the last time slot of each month (excluding the contract-ending month).
The term $\discount\cdot\mathbb{E}_t\left[V_{t+1}^{n}\left(\hat{q}^n_t,\dcap^{n},\bm{p}_{t+1} \right) \right]$ is the expected maximal  discounted payoff from slot $t+1$ to the end of the contract.
Here $\discount\in(0,1)$ is the time discount.
We use $\mathbb{E}_t[\cdot]$ to denote $\mathbb{E}_{x_n}\left[\mathbb{E}_{\bm{p}_{t+1}}[\cdot] \right]$ for brevity.\footnote{In practice, users only have prior beliefs of  the future trading prices instead of knowing their precise values in advance. The prior beliefs come from the observation of the past trading prices, with related discussions in \cite{villas2004consumer}.} 
Moreover, we have substituted (\ref{Equ: transition across}) in (\ref{Equ: problem last day of month}) here.

Case 3 corresponds to the time slots that  are not the last slot of any month.
We have substituted (\ref{Equ: transition inner}) in the third term of the RHS  of (\ref{Equ: problem middle}), i.e., $\discount\cdot\mathbb{E}_t\left[V_{t+1}^{n}\left(\hat{e}_{t}^{n},\hat{q}_{t}^{n} ,\bm{p}_{t+1} \right) \right] $.

%The third term inside the brackets, i.e., $\discount\cdot\mathbb{E}_t\left[V_{t+1}^{n}\left(\hat{e}_n,\hat{q}_n ,\bm{p}_{t+1} \right) \right]$, is the maximal expected discounted payoff from slot $t+1$ to the end of the period.

In each time slot $t$, user $n$ needs to make his optimal data trading decision $\bar{z}_t^{n*}$ based on the trading prices $\bm{p}_{t}$ and his leftover data volume $(e^n_t,q^n_t)$, while taking into account his random data demand $x^n_t$.
We will derive the users' optimal trading policy in Section \ref{Section: Trading Policy}.

\subsection{MNO's Decision}\label{Subsection: MNO Pricing}
We formulate MNO's pricing problem based on the above users' model.
Recall that user $n$ makes his trading decision $a_{t}^{n}$ based on his leftover data and the trading prices $\bm{p}_{t}=\{p_t^s,p_t^b\}$.
The user might become a seller (i.e., $a_{t}^{n}<0$) or a buyer (i.e., $a_{t}^{n}>0$), or choose not trade at all (i.e., $a_{t}^{n}=0$).
Therefore, the \textit{total trading market demand} (from all buyers)  is
\begin{equation}\label{Equ: total market demand}
D_t(\bm{p}_{t})=\sum\limits_{n\in\mathcal{N}}\left( a_{t}^{n} \right)^+,
\end{equation}
and the \textit{total trading market supply} (from all sellers) is 
\begin{equation}\label{Equ: total market supply}
S_t(\bm{p}_{t})=\sum\limits_{n\in\mathcal{N}}\left( -a_{t}^{n} \right)^+.
\end{equation}

In (\ref{Equ: total market demand}) and (\ref{Equ: total market supply}), all users' trading decision vector $\bm{a}_t=\{a_t^n,n\in\mathcal{N}\}$ depends on the prices $\bm{p}_{t}$.  
We will derive the corresponding more detailed expression in Section \ref{Section: MNO's Pricing} after analyzing users' optimal data trading policy in Section \ref{Section: Trading Policy}.

Given the total demand $D_t(\bm{p}_{t})$ and the total supply $S_t(\bm{p}_{t})$, the total transaction quantity in the market  becomes $\min\{D_t(\bm{p}_{t}),S_t(\bm{p}_{t})\}$.
The MNO obtains $p_t^b-p_t^s$ revenue from each unit of transaction data.
Therefore, we formulate the MNO's revenue-maximizing pricing problem as follows:
\begin{problem}[MNO's Pricing Problem]
	\begin{equation}
	\begin{aligned}
	\max\limits_{\bm{p}_{t}\ge\bm{0}}	&\  \left( p_{t}^{b}-p_{t}^{s} \right) \cdot \min \big\{S_t(\bm{p}_{t}), D_t(\bm{p}_{t}) \big\} .
	\end{aligned}
	\end{equation}
\end{problem}

Now we have introduced the full model.
Next we first study users' optimal data trading policy in Section \ref{Section: Trading Policy}, and then look at MNO's revenue-maximizing pricing in Section \ref{Section: MNO's Pricing}.

\section{Users' Trading Policy}\label{Section: Trading Policy}
Next we will study the user's optimal data trading policy under two different scenarios:
\begin{itemize} 
	\item \textit{Plain trading}: A user decides his trading action to maximize the total discounted payoff in the current month, if no rollover happens at the end of the current month.
	\item \textit{Rollover-involved trading}: A user decides his trading action to maximize the total discounted payoff in the future, if rollover happens at the end of the current month. 
\end{itemize}

Users will be  in the plain trading case if the MNO does not offer the rollover mechanism.
If the MNO offers, users will face the plain trading case in the last month of the contract period and the rollover-involved trading case in other months.

Next we study the trading policy under the two scenarios and then compare the key differences between them.
Since our analysis focuses on a generic user's optimal decision, we will suppress the superscript $n$ unless it is not clear.
Due to space limit, proofs are provided in the on-line technical report \cite{Technicalreport}.

\begin{figure*} 
\centering 
\setlength{\abovecaptionskip}{0pt}
\setlength{\belowcaptionskip}{0pt}
\subfigure[The $k$-th slot of the $M$-th month.]{\label{fig: Policy_pure}\includegraphics[width=0.3\linewidth]{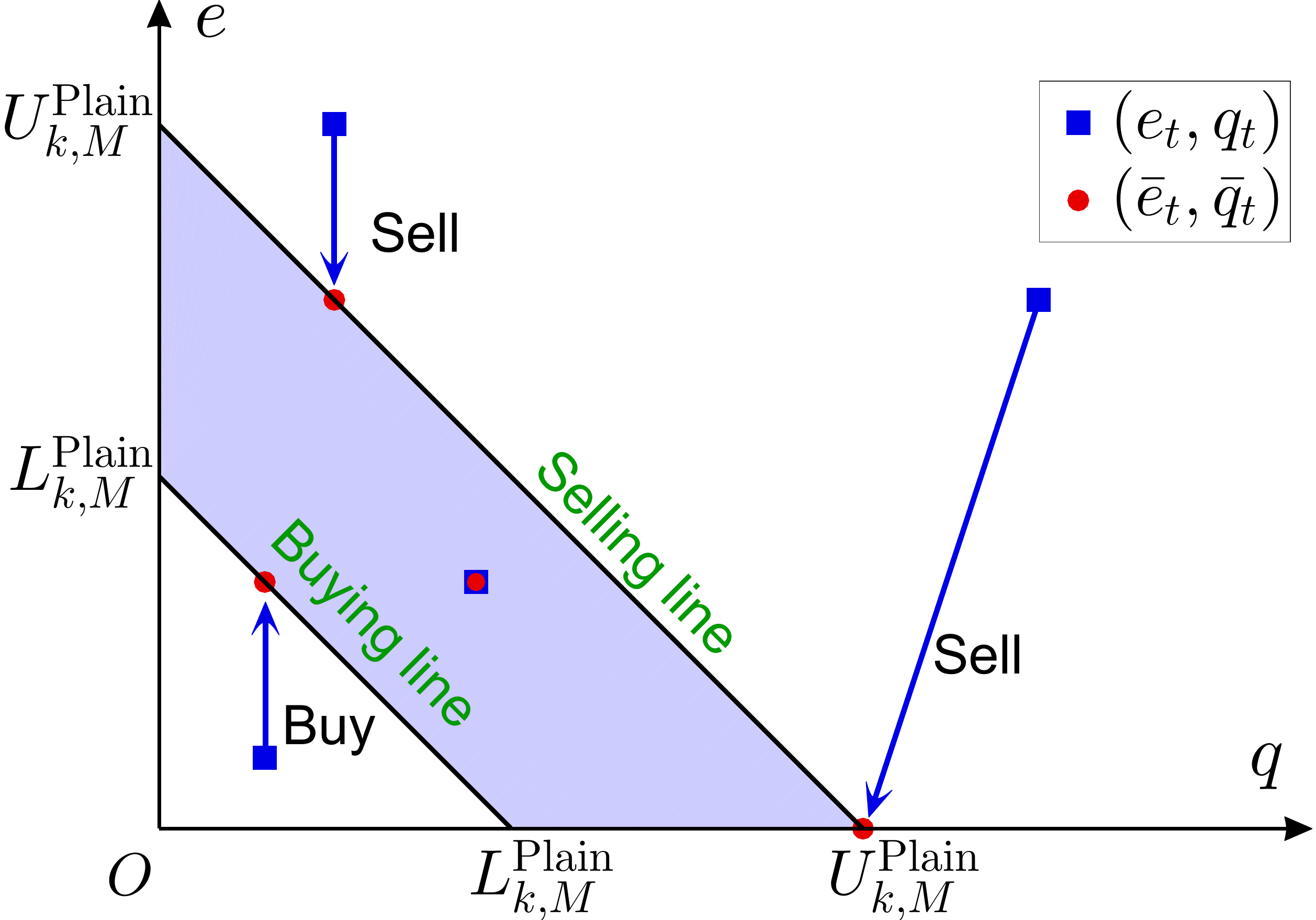}} \qquad
\subfigure[The $k$-th slot of the $m$-th month, $m<M$.]{\label{fig: Policy_rollover}\includegraphics[width=0.3\linewidth]{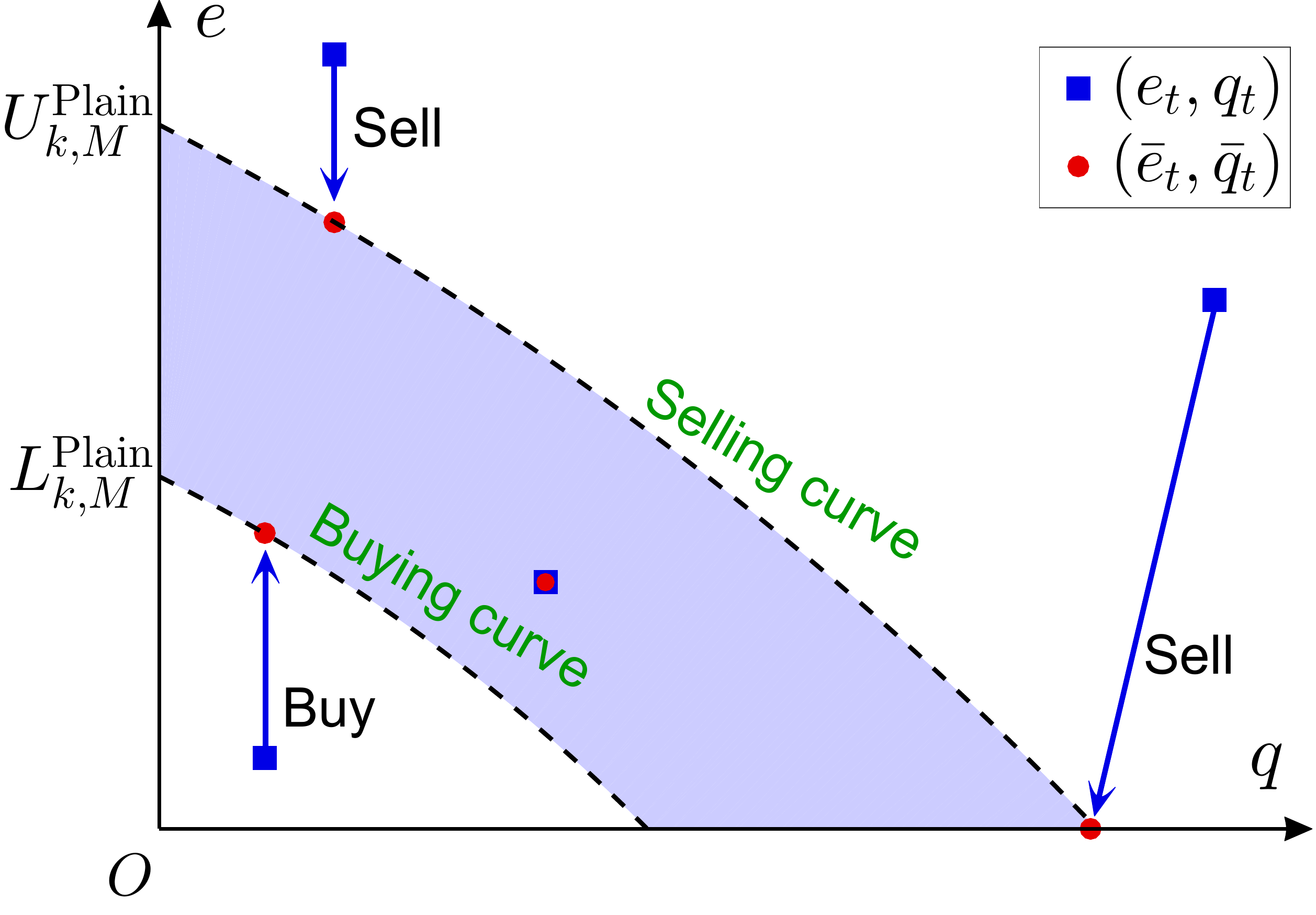}} \qquad
\subfigure[The same $k$-th slot of different months.]{\label{fig: Policy_compare}\includegraphics[width=0.3\linewidth]{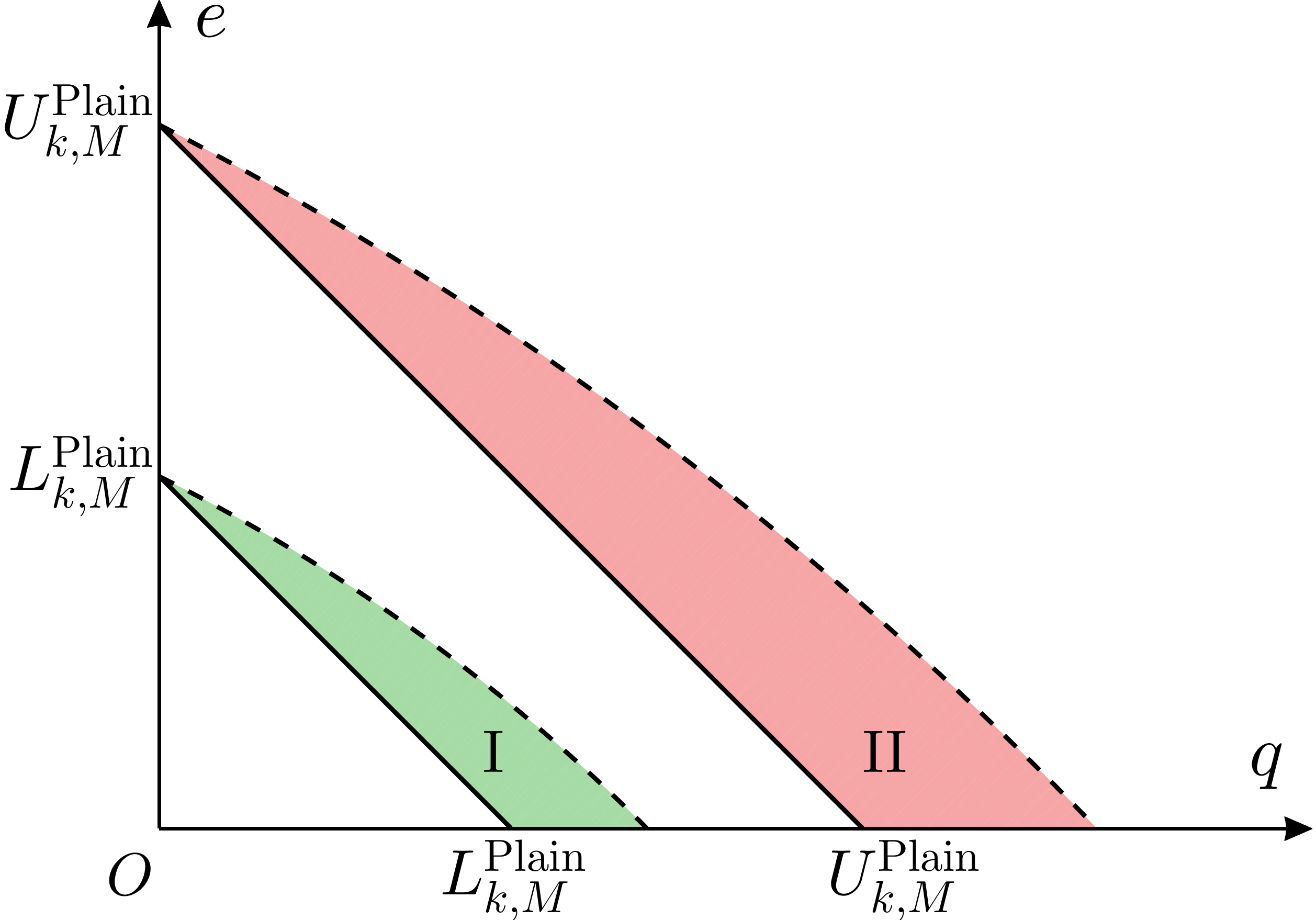}} 
\caption{Illustrations of the optimal trading policy given the trading price $\bm{p}=\{p^s,p^b\}$. \vspace{-5pt}
%(a) Plain Trading: the \textit{buying line} and \textit{selling line} correspond to $e+q=L^\text{Plain}_{k,M}(p^b)$ and $e+q=U^\text{Plain}_{k,M}(p^b)$. (b) Rollover-involved Trading: the \textit{buying curve} and \textit{selling curve} correspond to $e+q=L^\text{Roll}_{k,m}(p^b,q)$ and $e+q=U^\text{Roll}_{k,m}(p^s,q)$ for all time slots before the last month, i.e., $m<M$.
}
\end{figure*}

\subsection{Plain Trading}
Now we study the plain trading case.
Specifically, we will analyze the optimal trading policy of each time slots in the contract-ending ($M$-th) month, i.e., $t=K(M-1)+k$ and $k\in\{1,2,...,K\}$.
Theorem \ref{Theorem: trading policy of last month} summarizes the result. 
Due to page limit, we  provide proof sketches for Theorems \ref{Theorem: trading policy of last month} and \ref{Corollary: L U adfee early month},  and omit the proofs of other results.

\begin{theorem}[Plain Trading Policy]\label{Theorem: trading policy of last month}
	For any $t=K(M-1)+k$ and $k\in\{1,2,...,K\}$, given the leftover data volume  $(e,q)$ and the trading prices $\bm{p}=\{p^s,p^b\}$, there exists a pair of thresholds $\{L^\text{\rm Plain}_{k,M}(p^b),U^\text{\rm Plain}_{k,M}(p^s)\}$ with $L^\text{\rm Plain}_{k,M}(p^b)\le U^\text{\rm Plain}_{k,M}(p^s)$, such that the optimal trading action is 
	\begin{equation}
	\bar{z}_t^* =\left\{
	\begin{aligned}
	& L^\text{\rm Plain}_{k,M}(p^b),	\  \text{ if }\textstyle e+q< L^\text{\rm Plain}_{k,M}(p^b),\\
	& e+q,					\quad \ \ \  \text{ if }\textstyle L^\text{\rm Plain}_{k,M}(p^b) \le e+q \le U^\text{\rm Plain}_{k,M}(p^s),\\
	& U^\text{\rm Plain}_{k,M}(p^s),	\  \text{ if }\textstyle e+q > U^\text{\rm Plain}_{k,M}(p^s).
	\end{aligned}
	\right.
	\end{equation} 
\end{theorem}
%\begin{proof}[Proof Sketch]
%	Since the user's data trading problem (Problem \ref{Problem: Optimal Data Trading}) is a dynamic programming problem, we use backward induction to analyze the optimal policy based on the value functions \cite{bellman2013dynamic}.
%	Specifically, we start with $t=KM$ and derive the thresholds $\{L^\text{\rm Plain}_{K,M}(p^b),U^\text{\rm Plain}_{K,M}(p^s)\}$ according to
%	\begin{equation}
%	\left\{
%	\begin{aligned}
%		& F( L^\text{\rm Plain}_{K,M} )=1-{p^b}/{\adfee},\\
%		& F( U^\text{\rm Plain}_{K,M} )=1-{p^s}/{\adfee},
%	\end{aligned}
%	\right.		
%	\end{equation}	
%	where $F(\cdot)$ is the CDF of the user's daily data consumption.
%	
%	Then we proceed the induction on $t=K(M-1)+k$ (where $k=1,2,...,K-1$) for the following properties:
%	\begin{equation}
%	\left\{
%	\begin{aligned}
%	& G_{k,M}( L^\text{\rm Plain}_{k,M} )=1-{p^b}/{\adfee},\\
%	& G_{k,M}( U^\text{\rm Plain}_{k,M} )=1-{p^s}/{\adfee},\\
%	& L^\text{\rm Plain}_{k,M}\ge L^\text{\rm Plain}_{k+1,M},\ U^\text{\rm Plain}_{k,M}\ge U^\text{\rm Plain}_{k+1,M}.
%	\end{aligned}
%	\right.
%	\end{equation}
%	
%	Here $G_{k,M}(y)$ corresponds to a  series of intermediate functions, given by
%	\begin{equation}
%	G_{k,M}(\bar{z})\triangleq 
%	F(\bar{z})- \frac{\discount}{\adfee} \cdot \frac{\partial \mathbb{E}_t\left[V_{t+1} \left(\hat{e}_{t},\hat{q}_{t} ,\bm{p}_{t+1} \right) \right]}{\partial \bar{z}}  .
%	\end{equation}
%	
%%	Note that, in the above induction progress, we also derive the main properties of the two thresholds, which are going to be presented in Corollary \ref{Corollary: L U adfee last month}. 
%\end{proof}

Theorem \ref{Theorem: trading policy of last month} shows that the optimal plain trading policy is a \textit{target interval policy} specified by the buy-up-to threshold $L^\text{Plain}_{k,M}(p^b)$ and the sell-down-to threshold $U^\text{Plain}_{k,M}(p^s)$.
More specifically, if the user's total leftover data volume $e+q$ is less than $L^\text{Plain}_{k,M}(p^b)$, then the user needs to buy extra data and increase the leftover data volume to $L^\text{Plain}_{k,M}(p^b)$.
If the leftover data $e+q$ is higher than $U^\text{Plain}_{k,M}(p^s)$, then he needs to sell some data and reduce the leftover data volume to $U^\text{Plain}_{k,M}(p^s)$.
If the leftover data $e+q$ is already between these two thresholds, the user should choose not to trade. 

Fig. \ref{fig: Policy_pure} illustrates the optimal plain trading policy in the $k$-th time slot of the $M$-th month.
The horizontal and vertical axises correspond to the user's long-term data $q$ and the shot-term data $e$.
Each point in the plane specifies the data volume $(e,q)$.
The blue region between the \textit{selling line} and the \textit{buying line} represents those states where the user do not need to trade, i.e., $ L^\text{Plain}_{k,M}(p^b) \le e+q \le U^\text{Plain}_{k,M}(p^s)$.
The blue squares represent the  data volume before trading, i.e., $(e_t,q_t)$.
The red circles represent the data volume after trading, i.e., $(\bar{e}_t,\bar{q}_t)$.
We also represent the optimal trading action with the blue arrows.

%The thresholds $\{L^\text{Plain}_{k,M}(p^b),U^\text{Plain}_{k,M}(p^s)\}$ in Theorem \ref{Theorem: trading policy of last month} are derived from backward induction on $t$.
%{\color{red}Due to space limit, we provide more detailed explanations in  our technical report and focus on explaining the key insights in the main paper.}
Next we present the key properties of the threshold values in Corollary \ref{Corollary: L U adfee last month}.

%\begin{corollary}\label{Corollary: Buy infinite last month}
%	The two thresholds $L_t(p^b_t)$ and $U_t(p^s_t)$ go to positive infinity, if there exists an $i\ge t$ such that
%	\begin{equation}\label{Equ: Buy infinite last month}
%	\discount^i\mathbb{E}\left[p_{t-i}^s|\bm{p}_t\right] > p_t^b.
%	\end{equation} 	
%\end{corollary}
%The inequality condition  in Corollary \ref{Corollary: Buy infinite last month} specifies an extreme case.
%Specifically, $\discount^i\mathbb{E}\left[p_{t-i}^s|\bm{p}_t\right]$ in (\ref{Equ: Buy infinite last month}) is the discounted expected selling price at slot $t-i$.
%Intuitively, the user can benefits from buying infinite much data at slot $t$ and selling it out at slot $t-i$, if the buying price at slot $t$ is lower than the discounted expected selling price at slot $t-i$, i.e., $\discount^i\mathbb{E}\left[\adfee_{t-i}^s|\bm{p}_t\right]>p_t^b$.
%In practice, such a circumstance seldom happens, since the trading prices do not significantly change.\footnote{Take the data trading platform of CMHK as an example, the buying price is relatively stable and around HK$\$15$/GB.}
%In the subsequent analysis, we will focus on the key properties of the two thresholds and assume $\discount^i\mathbb{E}\left[\adfee_{t-i}^s|\bm{p}_t\right]\le p_t^b$.

%Recall that we have two time scales, i.e., $t=K(m-1)+k$ for $k$-th time slot of the $m$-th month.
%For notation convenience, we sometimes write the two thresholds as $L_{k,m}(\cdot)$ and $U_{k,m}(\cdot)$  unless confusion would otherwise arise.

\begin{corollary}\label{Corollary: L U adfee last month}
	The two thresholds $\{L^\text{\rm Plain}_{k,M}(p^b),U^\text{\rm Plain}_{k,M}(p^s)\}$ in Theorem \ref{Theorem: trading policy of last month} have  the following properties.
	\begin{enumerate} 
		\item The buy-up-to threshold $L^\text{\rm Plain}_{k,M}(p^b)$ decreases in $p^b$.
		\item The sell-down-to threshold $U^\text{\rm Plain}_{k,M}(p^s)$ decreases in $p^s$.
		\item Given the trading price $\bm{p}$, we have
		\begin{equation}
		\left\{
		\begin{aligned}
		& L^\text{\rm Plain}_{1,M}(p^b)\ge L^\text{\rm Plain}_{2,M}(p^b)\ge...\ge L^\text{\rm Plain}_{K,M}(p^b),\\
		& U^\text{\rm Plain}_{1,M}(p^s)\ge U^\text{\rm Plain}_{2,M}(p^s)\ge...\ge U^\text{\rm Plain}_{K,M}(p^s).
		\end{aligned}
		\right.
		\end{equation}
	\end{enumerate}
\end{corollary}

In Corollary \ref{Corollary: L U adfee last month}, the first property indicates that a higher buying price $p^b$ leads to a lower buy-up-to threshold $L^\text{Plain}_{k,M}(p^b)$, hence the user tends to buy less data.
The second property indicates that a higher selling price $p^s$ leads to a lower sell-down-to threshold $U^\text{Plain}_{k,M}(p^s)$, hence the user will sell more data.
The third property shows that the thresholds decrease in time.
This is because that the need to maintain a high data inventory decreases over time.
%These properties not only hold for the plain trading case.

%Next let us move on to the rollover-involved trading.

\subsection{Rollover-involved Trading}
Now we study the rollover-involved trading and analyze the optimal trading policy in each time slot before the last month, i.e., $t=K(m-1)+k$ where $k\le K$ and $m<M$.
We first present the optimal rollover-involved trading policy in Theorem \ref{Theorem: trading policy of early month} and then elaborate it in details.
\begin{theorem}[Rollover-involved Trading Policy]\label{Theorem: trading policy of early month}
	For all $t=K(m-1)+k$ and $m<M$, given the data volume $(e,q)$ and the trading prices $\bm{p}=\{p^s,p^b\}$, there exists a pair of thresholds $\{L^\text{\rm Roll}_{k,m}(p^b,q),U^\text{\rm Roll}_{k,m}(p^s,q)\}$ with $L^\text{\rm Roll}_{k,m}(p^b,q)\le U^\text{\rm Roll}_{k,m}(p^s,q)$, such that the optimal trading action is
	\begin{equation}
		\bar{z}_t^*=\left\{
		\begin{aligned}
			& L^\text{\rm Roll}_{k,m}(p^b,q),		& \text{if }&\textstyle e+q< L^\text{\rm Roll}_{k,m}(p^b,q),\\
			& e+q,								& \text{if }&\textstyle L^\text{\rm Roll}_{k,m}(p^b,q) \le e+q \le U^\text{\rm Roll}_{k,m}(p^s,q),\\
			& U^\text{\rm Roll}_{k,m}(p^s,q),		& \text{if }&\textstyle e+q > U^\text{\rm Roll}_{k,m}(p^s,q).
		\end{aligned}
		\right.
	\end{equation} 	
\end{theorem}

%\begin{proof}[Proof Sketch]
%	The proof of Theorem \ref{Theorem: trading policy of early month} follows a similar procedure as Theorem \ref{Theorem: trading policy of last month}.
%	The difference exists in the time slot $t=K(M-1)$, where the user need to decide his trading policy according to (\ref{Equ: problem last day of month}) instead of (\ref{Equ: problem last day of contract}).
%\end{proof}

Theorem \ref{Theorem: trading policy of early month} shows that the optimal rollover-involved trading policy  is still a \textit{target interval policy}, which is similar to that in the plain trading case.
However, the buy-up-to threshold $L^\text{Roll}_{k,m}(p^b,q)$ and the sell-down-to threshold $U^\text{Roll}_{k,m}(p^b,q)$ not only depends on the trading price $\bm{p}$, but also the leftover long-term data volume $q$.
This is because that the long-term data $q$ also plays a role in the next month.
Before we illustrate the rollover-involved trading policy, let us first introduce some key properties in Corollary \ref{Corollary: L U adfee early month}.

\begin{corollary}\label{Corollary: L U adfee early month}
	For all $t=K(m-1)+k$ and $m<M$, the thresholds $\{L^\text{\rm Roll}_{k,m}(p^b,q),U^\text{\rm Roll}_{k,m}(p^s,q)\}$ in Theorem \ref{Theorem: trading policy of early month} have the following properties:
	\begin{enumerate} 
		\item The buy-up-to threshold $L^\text{\rm Roll}_{k,m}(p^b,q)$ decreases in $p^b$.
		\item The sell-down-to threshold $U^\text{\rm Roll}_{k,m}(p^s,q)$ decreases in $p^s$.
		\item Given the trading prices $\bm{p}$, we have 
		\begin{equation}
		\left\{
		\begin{aligned}
		& L^\text{\rm Roll}_{1,m}(p^b,q)\ge L^\text{\rm Roll}_{2,m}(p^b,q)\ge...\ge L^\text{\rm Roll}_{K,m}(p^b,q),\\
		& U^\text{\rm Roll}_{1,m}(p^s,q)\ge U^\text{\rm Roll}_{2,m}(p^s,q)\ge...\ge U^\text{\rm Roll}_{K,m}(p^s,q).
		\end{aligned}
		\right.
		\end{equation}
		\item If $q=0$, then we have
		\begin{equation} \label{Equ: L U q=0}
		\left\{
		\begin{aligned}
		& L^\text{\rm Roll}_{k,m}(p^b,0)=L^\text{\rm Plain}_{k,M}(p^b),\\
		& U^\text{\rm Roll}_{k,m}(p^s,0)=U^\text{\rm Plain}_{k,M}(p^s).
		\end{aligned}
		\right.
		\end{equation}
		
	\end{enumerate}
\end{corollary}

The first three properties of Corollary \ref{Corollary: L U adfee early month} indicate similar intuitions as those  in Corollary \ref{Corollary: L U adfee last month}.
The fourth property of Corollary \ref{Corollary: L U adfee early month} further shows that the two thresholds of the rollover-involved case degenerate into those of plain trading if there is no long-term data, i.e., $q=0$.
That is, the plain trading policy is a special case of the rollover-involved trading policy.

Fig. \ref{fig: Policy_rollover} illustrates the rollover-involved trading policy in the $k$-th time slot of the $m$-th month, where $m<M$.
Here the blue region between the \textit{selling curve} and the \textit{buying curve} represents those states where the user does not need to trade, i.e., $L^\text{\rm Roll}_{k,m}(p^b,q) \le e+q \le U^\text{\rm Roll}_{k,m}(p^s,q)$.
By comparing Fig. \ref{fig: Policy_rollover} with Fig. \ref{fig: Policy_pure}, we note that the rollover mechanism changes the straight selling and buying lines (as in Fig. \ref{fig: Policy_pure}) into nonlinear curves (as in Fig. \ref{fig: Policy_rollover}).
In the following, we discuss more insights on the trading thresholds and examine the impact of rollover mechanism.

\subsection{Impact of Rollover Mechanism}\label{Subsection: Impact of Rollover Mechanism}
Based on the analysis of the optimal trading policy of the two cases, we investigate the impact of rollover mechanism on users' trading behaviors in Corollary \ref{Corollary: rollover on trading}.
\begin{corollary}\label{Corollary: rollover on trading}
	Considering the same $k$-th time slot of different months, given the trading prices $\bm{p}=\{p^s,p^b\}$, the corresponding thresholds satisfy 
	\begin{equation}
		\left\{
		\begin{aligned}
			& L^\text{\rm Roll}_{k,1}(p^b,q) \ge L^\text{\rm Roll}_{k,2}(p^b,q) \ge...\ge L^\text{\rm Plain}_{k,M}(p^b),\\
			& U^\text{\rm Roll}_{k,1}(p^s,q) \ge U^\text{\rm Roll}_{k,2}(p^s,q) \ge...\ge U^\text{\rm Plain}_{k,M}(p^s).
		\end{aligned}
		\right. 
	\end{equation}
\end{corollary}

We illustrate Corollary \ref{Corollary: rollover on trading} by combining Fig. \ref{fig: Policy_pure} and Fig. \ref{fig: Policy_rollover} together in Fig. \ref{fig: Policy_compare}.
%By comparing the buying curve (or line) and selling curve (or line) in the $k$-th time slot of $M$-th month and $m$-th month ($m<M$), 
We note that there are two regions (i.e., Regions I and II) indicating different trading actions:
\begin{itemize} 
	\item Region I: The optimal policy is buying data based on the dash curves (with rollover) , but no-trading based on the solid lines (plain trading).
	That's, the rollover mechanism makes  users buy more data.
	\item Region II: The optimal policy is no-trading based on the dash curves (with rollover), but selling data based on the solid lines (plain trading).
	That's, the rollover mechanism makes users sell less data.
\end{itemize}

To sum up, we find that the rollover mechanism makes users hold more data by shifting the buying curve and selling curve upwards.
A high data inventory not only helps reduce the potential overage charge, but also increases the potential selling income (as the user can wait for higher selling prices).

%Now we have understood users' optimal trading behaviors.
%Next let us further look at the MNO's pricing problem.

\section{MNO's Pricing}\label{Section: MNO's Pricing}
We study the MNO's revenue-maximizing pricing problem, considering users' optimal data trading policy derived in Theorems \ref{Theorem: trading policy of last month} and \ref{Theorem: trading policy of early month}.
In this section, we will recover the superscript $n$ for each user.

Given all users' leftover data volume $\bm{e}_t$ and $\bm{q}_t$ in time slot $t$, under the trading prices $\bm{p}_t=\{p^s_t,p^b_t\}$, the \textit{total trading market demand} (from buyers) is given by
\begin{equation}\label{Equ: total market demand LU}
D_t(p^b_t)=\sum\limits_{n\in\mathcal{N}}\big( L^n_t(p^b_t,q^n_t)-e_t^n-q_t^n \big)^+,
\end{equation}
where $L^n_t(p^b_t,q^n_t)$ is the buy-up-to threshold of user $n$ in time slot $t$.

Similarly, the \textit{total trading market supply} (from sellers) is 
\begin{equation}\label{Equ: total market supply LU}
S_t(p^s_t)=\sum\limits_{n\in\mathcal{N}}\big( e_t^n+q_t^n-U^n_t(p^s_t,q^n_t) \big)^+ ,
\end{equation}
where $U^n_t(p^s_t,q^n_t)$ is the sell-down-to threshold of user $n$ in time slot $t$.

%Before we present the MNO's optimal pricing, let us first introduce the concept of \textit{perfect competition}.
%In economics, {perfect competition} is defined by several conditions (see \cite{frank1991microeconomics} for more details).
%One of these conditions is that \textit{no participant has the market power to set prices}.
%In our problem, the data trading market is not perfectly competitive, since the MNO determines the prices.
%In addition, it has been theoretically demonstrated that a perfectly competitive market will reach an equilibrium where the quantity supplied equals the quantity demanded \cite{frank1991microeconomics}.
%Next we show the somewhat surprising result that the MNO's revenue-maximizing pricing still clears the market, even though the data trading market does not exhibit the perfect competition.

\begin{figure}
	\centering 
	\subfigure[Demand $>$ Supply]{\label{fig: MNOPricing_2}\includegraphics[height=0.34\linewidth]{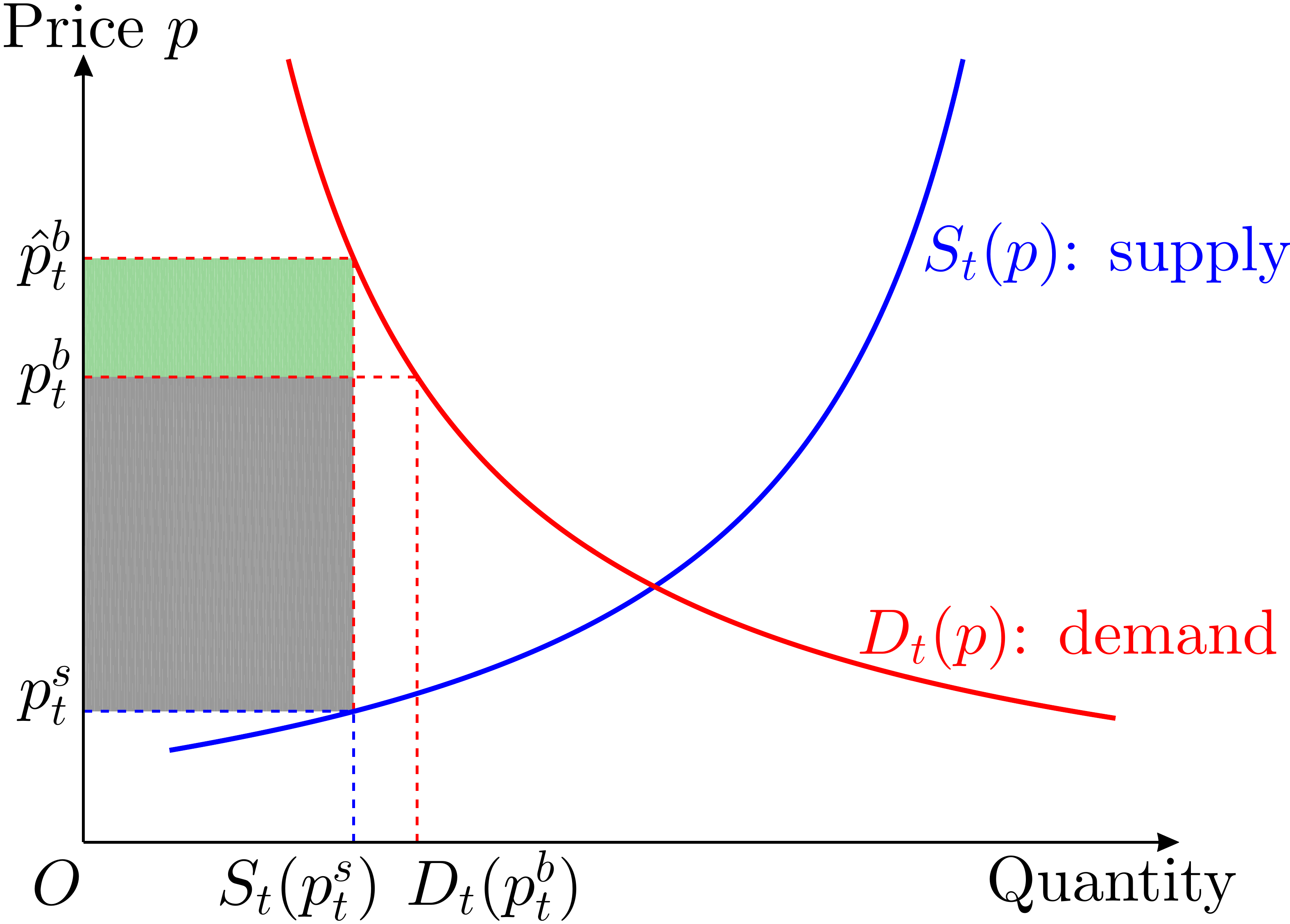}} \quad
	\subfigure[Demand $<$ Supply]{\label{fig: MNOPricing_1}\includegraphics[height=0.34\linewidth]{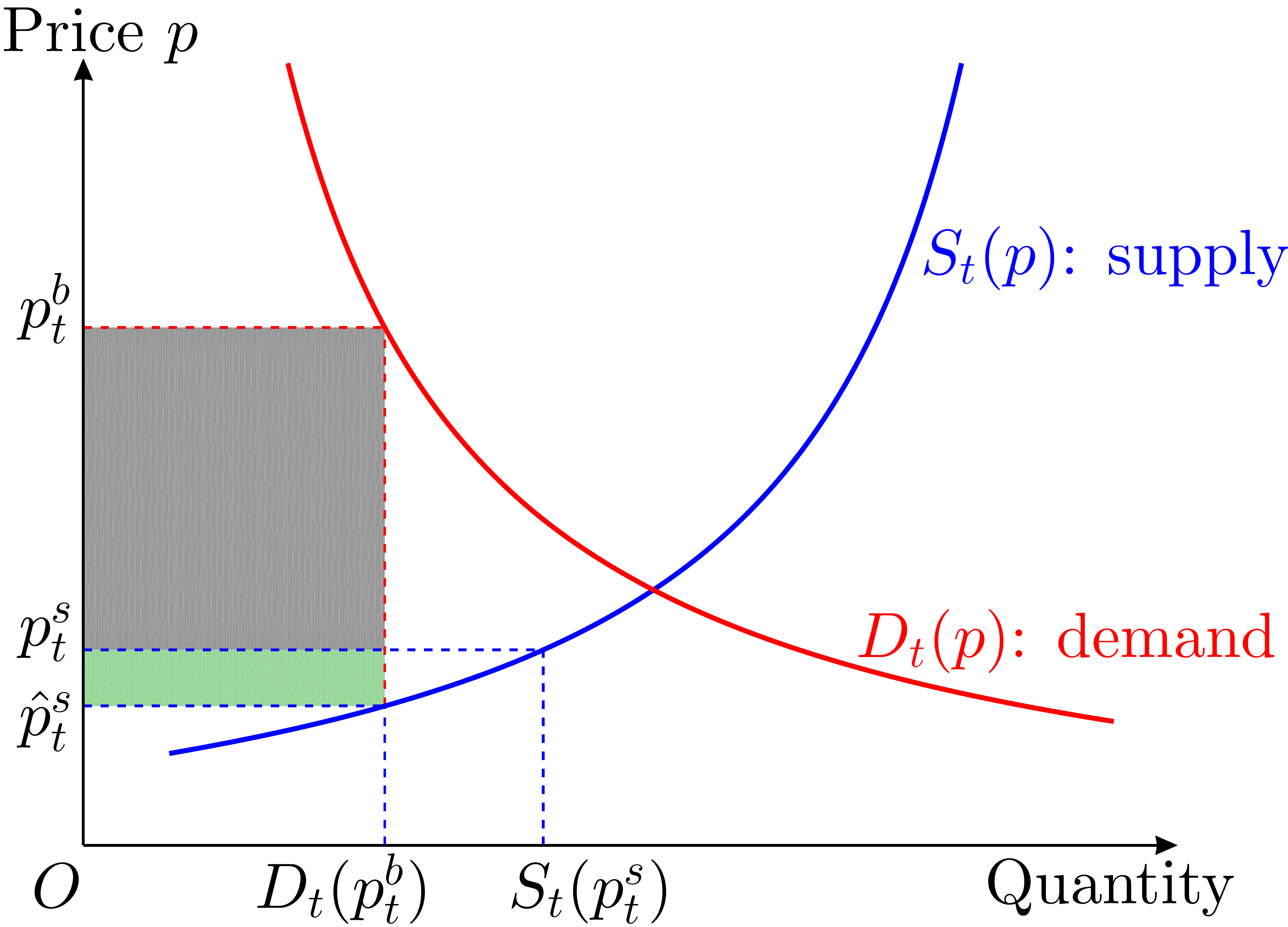}} 
	\caption{An illustration of the market clear pricing.\vspace{-10pt}}
	\label{fig: market clear}
\end{figure}

Fig. \ref{fig: market clear}  illustrates the different scenarios of demand and supply.
In each sub-figure, the vertical and horizontal axises correspond to the price (labeled by $p$) and the quantity, respectively.
The two curves are the demand curve $D_t(p)$ and the supply curve $S_t(p)$.
\begin{itemize} 
	\item Fig. \ref{fig: MNOPricing_2}: If the total demand is larger than the total supply, i.e.,  $S_t(p^s_t)<D_t(p^b_t)$, then the area of gray region is the MNO's total revenue, i.e., $(p^b_t-p^s_t)\cdot S_t(p^s_t)$.
	It is obvious that the MNO can increase its revenue by raising the buying price from $p^b_t$ to $\hat{p}^b_t$.
	Accordingly, the area of green region is the MNO's revenue increment.
	\item Fig. \ref{fig: MNOPricing_1}: If the total demand is smaller than the total supply, i.e.,  $S_t(p^s_t)>D_t(p^b_t)$, then it is obvious that the MNO can increases its revenue by decreasing the selling price from $p^s_t$ to $\hat{p}^s_t$.	
	Similarly, the area of green region is the revenue increment for the MNO.
\end{itemize}

Based on the above insights, we can conclude that the MNO should set the price such that the demand equals to the supply in the data trading market.
Next we characterize the MNO's optimal prices in Theorem \ref{Theorem: MNOPricing}.
For notation simplicity, we first define $P_{t,D}(\theta)$ and $P_{t,S}(\theta)$ as the inverse functions of $D_t(p)$ and $S_t(p)$, respectively.
Here $\theta$ is the quantity of total market demand or supply.

\begin{theorem}\label{Theorem: MNOPricing}
	In time slot $t$, the MNO's revenue-maximizing prices, denoted by $\{\tilde{p}_t^s,\tilde{p}_t^b\}$, are given by
	\begin{equation}
	\left\{
	\begin{aligned}
		& \tilde{p}_t^s = P_{t,S}({\theta}^*), \\
		& \tilde{p}_t^b = P_{t,D}({\theta}^*),
	\end{aligned}
	\right.
	\end{equation} 
	where ${\theta}^*$ is the optimal transaction quantity, given by
	\begin{equation}
	\begin{aligned}
		P_{t,D}({\theta}^*) + {\theta}^*\cdot P_{t,D}'(\theta^*) =   P_{t,S}(\theta^*) + {\theta}^*\cdot P_{t,S}'({\theta}^*).
	\end{aligned}
	\end{equation}
	
%	Here both $P_D'({\theta},\bm{e}_t,\bm{q}_t)$ and $P_S'({\theta},\bm{e}_t,\bm{q}_t)$ are the partial derivative regarding $\theta$.
\end{theorem}

In Theorem \ref{Theorem: MNOPricing}, we characterize the optimal selling price $\tilde{p}_t^s$ and buying price $\tilde{p}_t^b$ through the optimal transaction quantity $\theta^*$ based on the demand and supply curves.
In practice, the MNO can estimate the demand and supply curve based on all users' leftover data volumes $\bm{e}_t$ and $\bm{q}_t$.
In Section \ref{Section: numerical results}, we will further quantitatively evaluate the MNO's revenue based on the real-world data.

\begin{figure} 
	\centering 
		\setlength{\abovecaptionskip}{1pt}
		\setlength{\belowcaptionskip}{0pt}
	\subfigure[User 1: $\mu=15.2$, $\sigma=11.5$]{\label{fig: Zhiyuan_pdf}\includegraphics[width=0.48\linewidth]{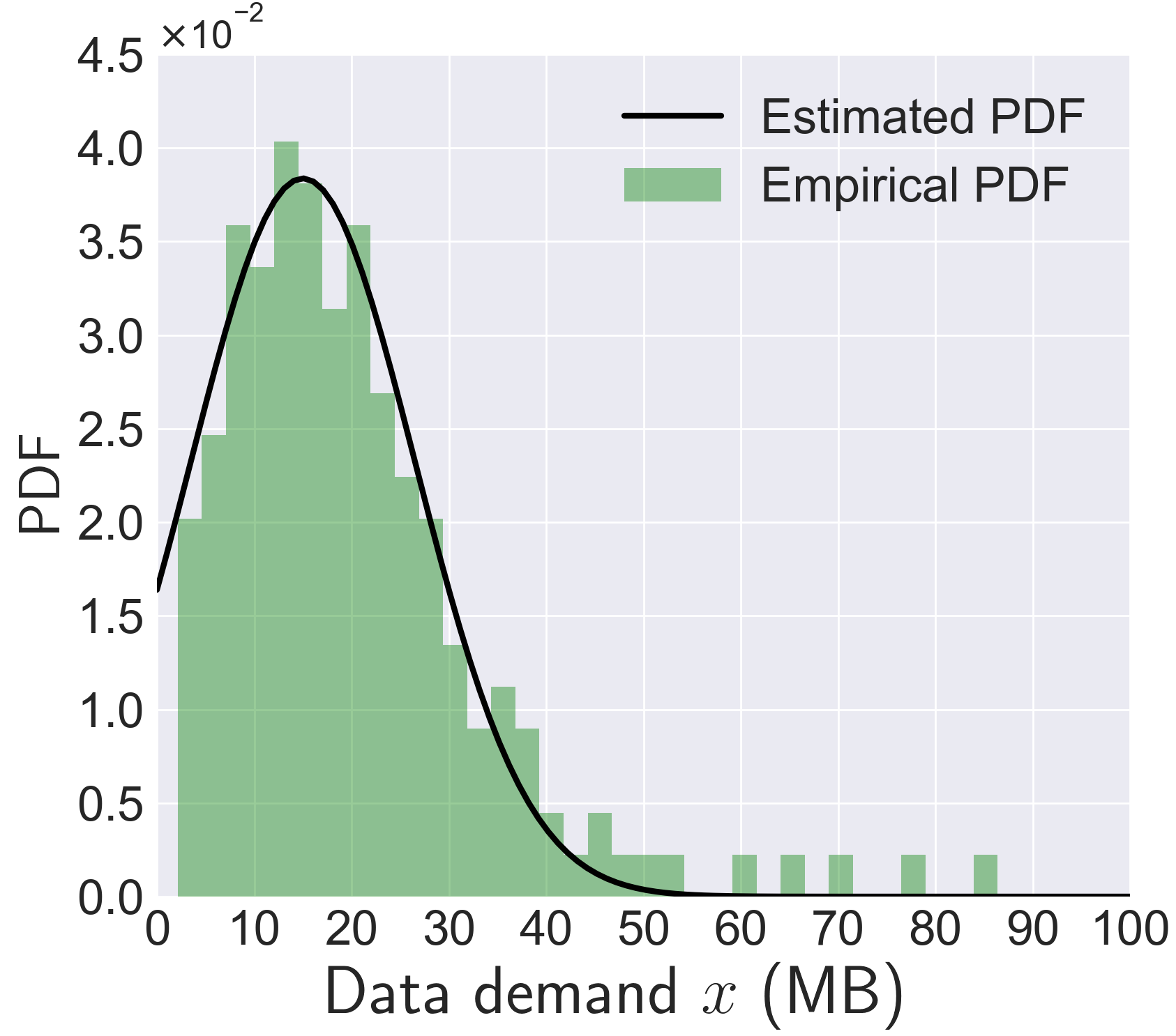}}  \
	\subfigure[User 2: $\mu=70.2$, $\sigma=46.1$]{\label{fig: Jiang_pdf}\includegraphics[width=0.48\linewidth]{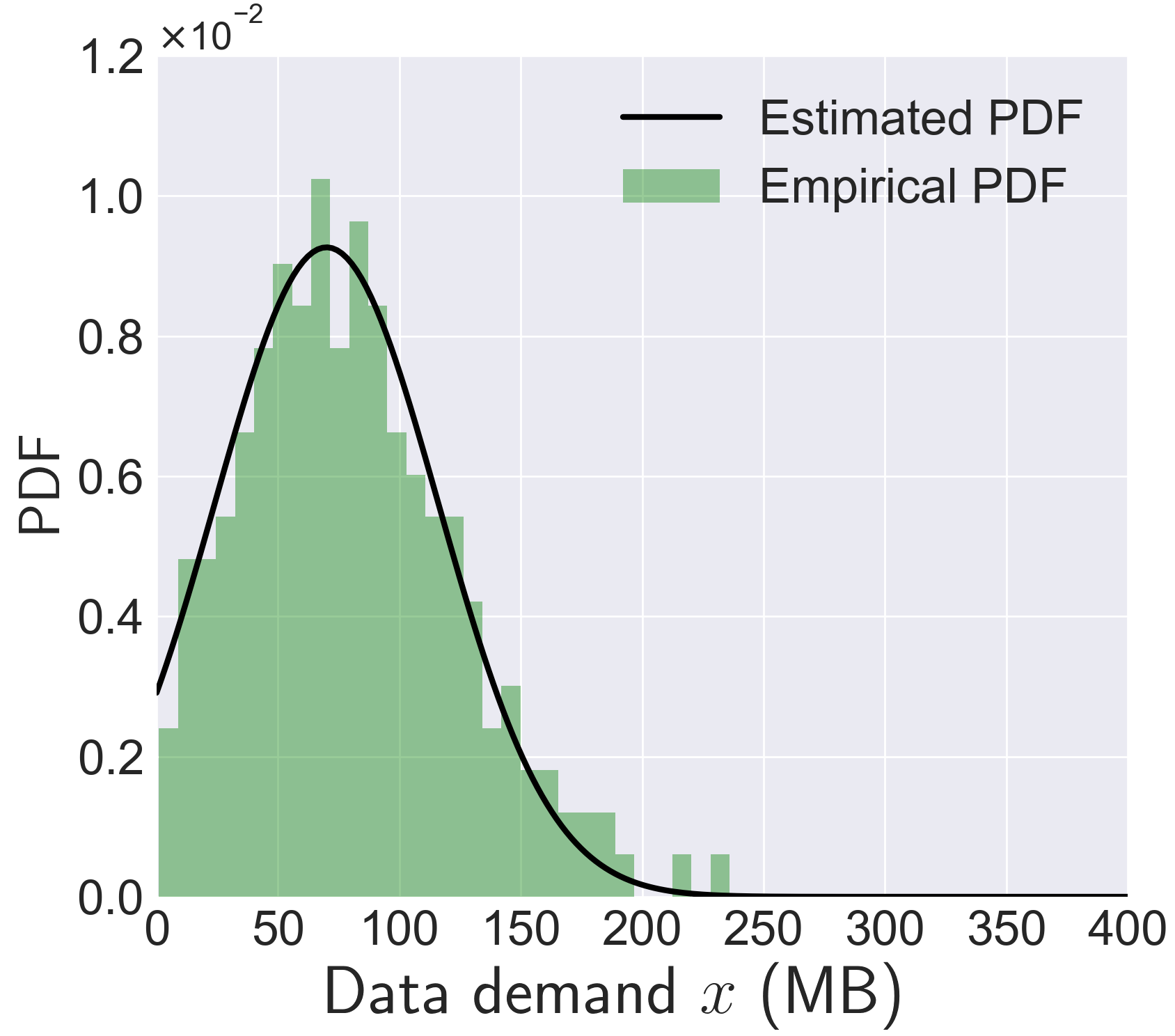}} 
	\caption{Fitting results based on two users' empirical data.\vspace{-10pt}}
	%		The corresponding $p$-value for the users are $1\times10^{-3}$ and $1.1\times10^{-3}$.	
	\label{fig: fitting}
\end{figure}

\section{Numerical Results}\label{Section: numerical results}
We apply our analysis to a real-world usage trace.
We collect a group of mobile users' data consumption records in China from December 2017 to June 2018.
For each user in our dataset, we have the information of the data consumption and the corresponding time duration for each Internet connection.\footnote{Mobile users in China can download their data consumption trace of the previous six months. We collect the data from the volunteered mobile users. We refer interested readers to our technical report for more details \cite{Technicalreport}.} 
We first use the empirical data to estimate users' demand distributions, then compare the buy-up-to and sell-down-to thresholds based on the data.
Finally, we evaluate the impact of rollover mechanism on MNO revenue and user payoffs.

\subsection{Empirical Results}
Mobile users' data consumption highly depends on their daily activity and mobility.
Many statistical studies (e.g., \cite{InternetUsage}) have shown a clear periodic nature for Internet connections and the period is twenty-four hours.
Therefore, we follow the previous studies in \cite{Zheng2018dynamics} by viewing one day as the minimum time slot
and estimate users' daily demand distribution by assuming a normal distribution $N(\mu,\sigma^2)$ truncated at zero.
%The corresponding probability density function (PDF) is given by 
%\begin{equation}
%f(x;\mu,\sigma^2)=\frac{ \phi\left( \frac{x-\mu}{\sigma} \right) }{ \sigma \left[1-\Phi\left(-\frac{\mu}{\sigma}\right)\right] },\quad\ \forall\ x\ge0,
%\end{equation}
%where $\phi(\cdot)$ and $\Phi(\cdot)$ are the PDF and CDF of the standard normal distribution, respectively.
The mean $\mu$ and the standard deviation $\sigma$ are the two parameters that we need to estimate for each individual.
To proceed, we first estimate the two parameters to minimize the least squares divergence between the estimated and empirical PDFs, then verify the goodness-of-fit statistically using Kolmogorov-Smirnov test.

Due to the space limit, here we take two users as examples.
In Fig. \ref{fig: fitting}, we plot the fitting results of the low-demand User 1 (with a $0.5$GB data cap) and the high-demand User 2 (with a $2$GB data cap).
Here the green bars and the black curves are the empirical PDFs and estimated PDFs, respectively.
The mean $\mu$ and standard derivation $\sigma$ are labeled in the captions. 
%We can see that the curves fits well in each case. 
%In fact, using the Kolmogorov-Smirnov test \cite{massey1951kolmogorov}, we have also statistically verified the goodness-of-fit and show the corresponding $p$-value in the captions.
%To further investigate the goodness-of-fit statistically, we conduct the Kolmogorov-Smirnov test on the null hypothesis that the daily demand come from the estimated distribution and show the corresponding $p$-value (at the significance level of $p$ we cannot rejects the hypothesis \cite{massey1951kolmogorov}) in the captions.

\subsection{Optimal Trading Policy}
Next we simulate the trading thresholds for each user based on the estimated data consumption distribution.
To make reasonable comparison, here we fix the trading prices as $p^s=10$HKD/GB and $p^b=15$HKD/GB. 
We will examine the time-variant prices in Section \ref{Subsection: Performance Evaluation}.
%using the market price of the trading platform of CMHK.

%We record the daily transaction price in June 2018 and find that the trading prices are rather stable, with $p^s=10$HKD/GB and $p^b=15$HKD/GB. 
%Therefore, we use these price values in the following simulation.

\subsubsection{Plain Trading}
%We first simulate the plain trading scenario.
%As mentioned in Theorem \ref{Theorem: trading policy of last month}, the buy-up-to threshold $L^\text{Plain}_{k,M}(p^b)$ and the sell-down-to threshold $U^\text{Plain}_{k,M}(p^s)$ only depend on the trading prices $\bm{p}$.

\begin{figure}
		\setlength{\abovecaptionskip}{1pt}
		\setlength{\belowcaptionskip}{0pt}
	\centering 
	\subfigure[User 1]{\label{fig: Threshold_Pure_User1}\includegraphics[height=0.42\linewidth]{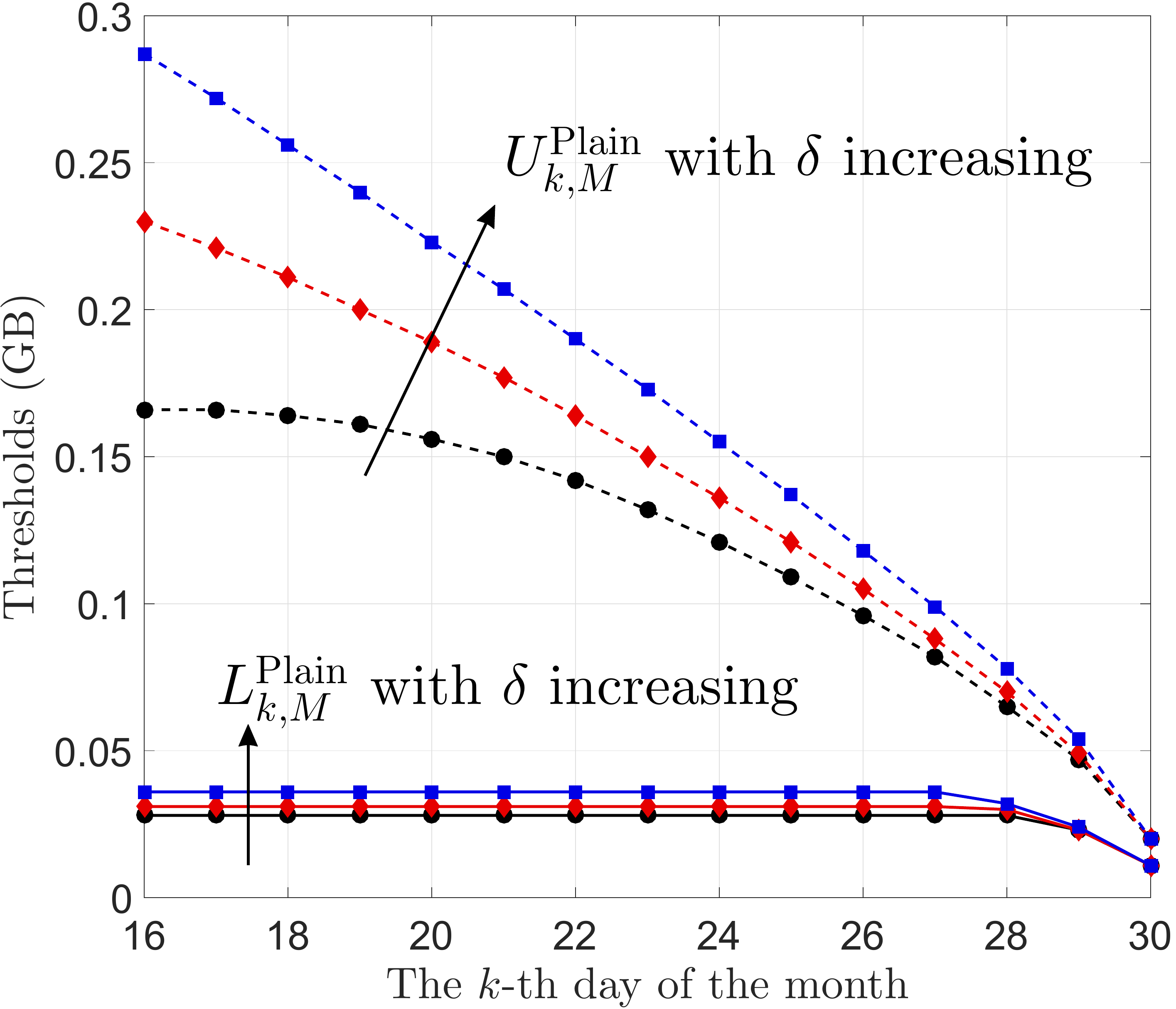}}  \
	\subfigure[User 2]{\label{fig: Threshold_Pure_User3}\includegraphics[height=0.42\linewidth]{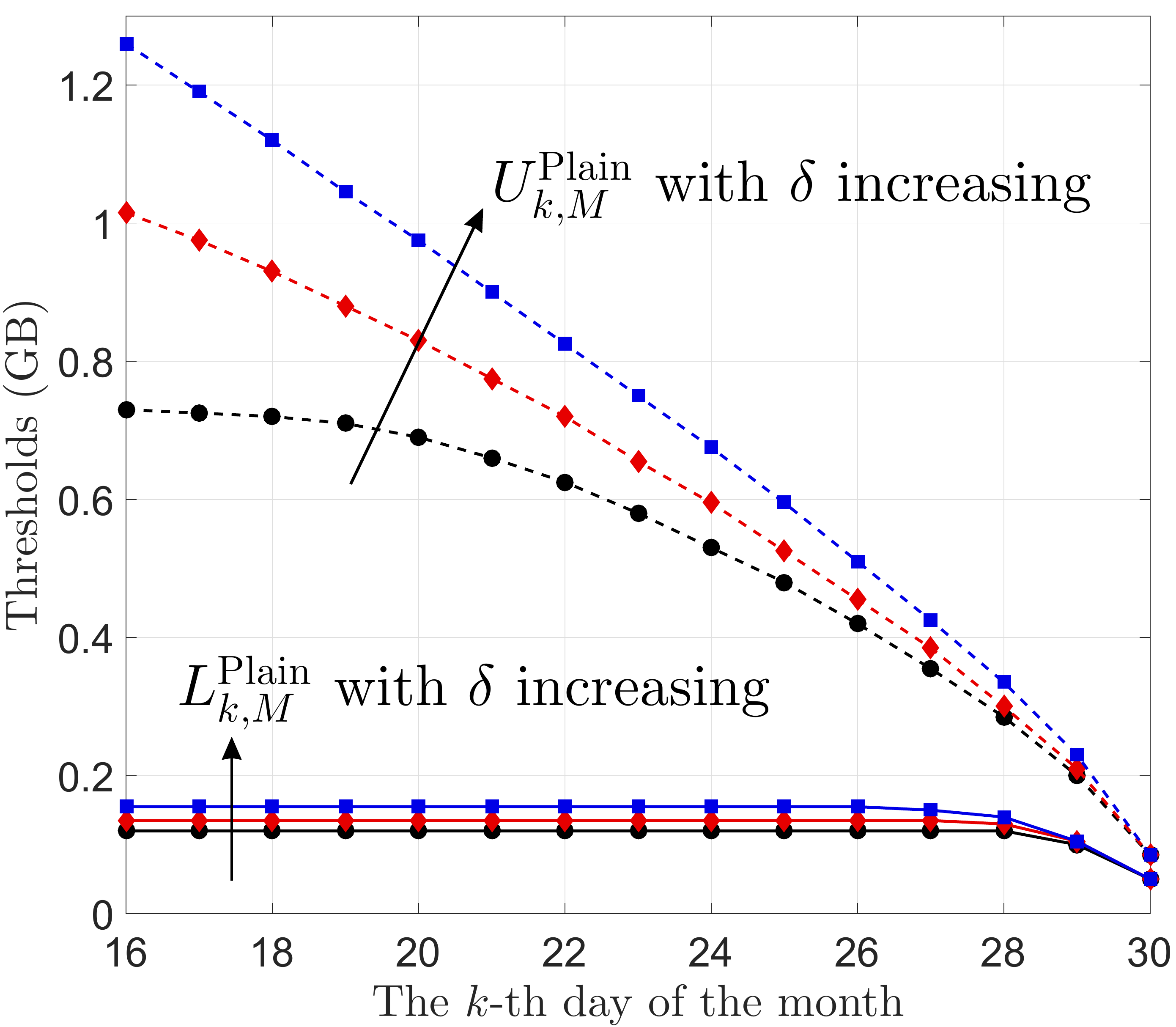}} 
	\caption{Plain trading thresholds of User 1 and User 2 under different time discounts $\discount\in\{0.92,0.95,0.98\}$. }
	\label{fig: Threshold_Pure}
\end{figure}

Fig. \ref{fig: Threshold_Pure} shows the trading thresholds of User 1 and User 2 in the plain trading case.
In each sub-figure, the horizontal axis represents the $k$-th day of the month.
For the illustration purpose, we only show the results of the second half month, i.e., $k=16,17,...,30$.
Moreover, we investigate the impact of time preference by considering three time discount values, i.e., $\discount\in\{0.92,0.95,0.98\}$.

Overall, both of the buy-up-to threshold and sell-down-to threshold decrease in $k$.
However, the value of buy-up-to threshold $L^\text{Plain}_{k,M}$ is less sensitive to $k$ and always remain small. 
This is because that users do not need to buy a lot of extra data in advance.
Instead, they prefer to maintaining a small amount of leftover data to avoid overage fee in the current day, then only buy more data when the data consumption reaches a significant level in the current month. 

\begin{figure}
	\centering
		\setlength{\abovecaptionskip}{1pt}			
		\setlength{\belowcaptionskip}{0pt}
	\subfigure[User 1]{\label{fig: Threshold_Roll_range_User1}\includegraphics[height=0.42\linewidth]{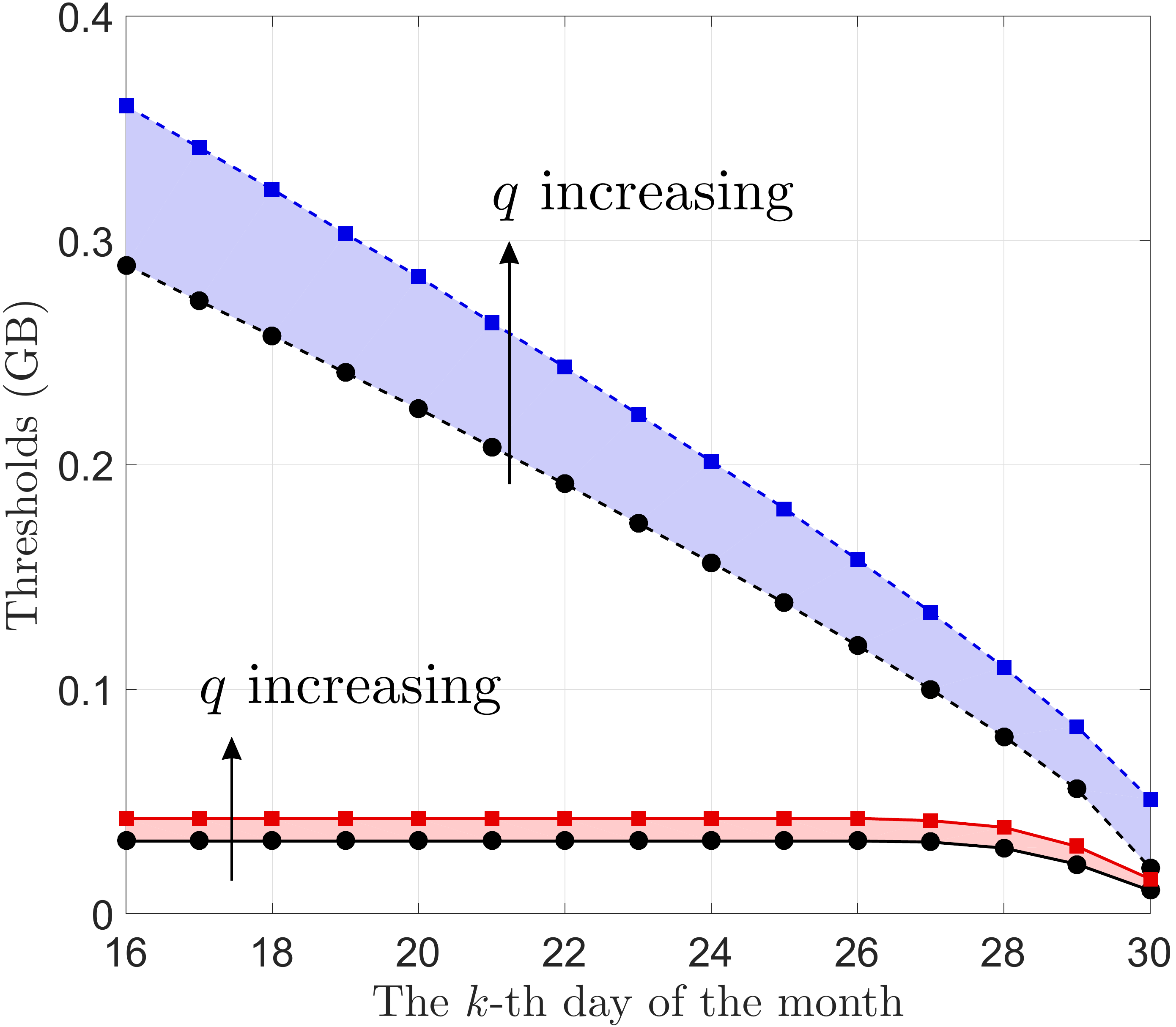}} \  
	\subfigure[User 2]{\label{fig: Threshold_Roll_range_User3}\includegraphics[height=0.42\linewidth]{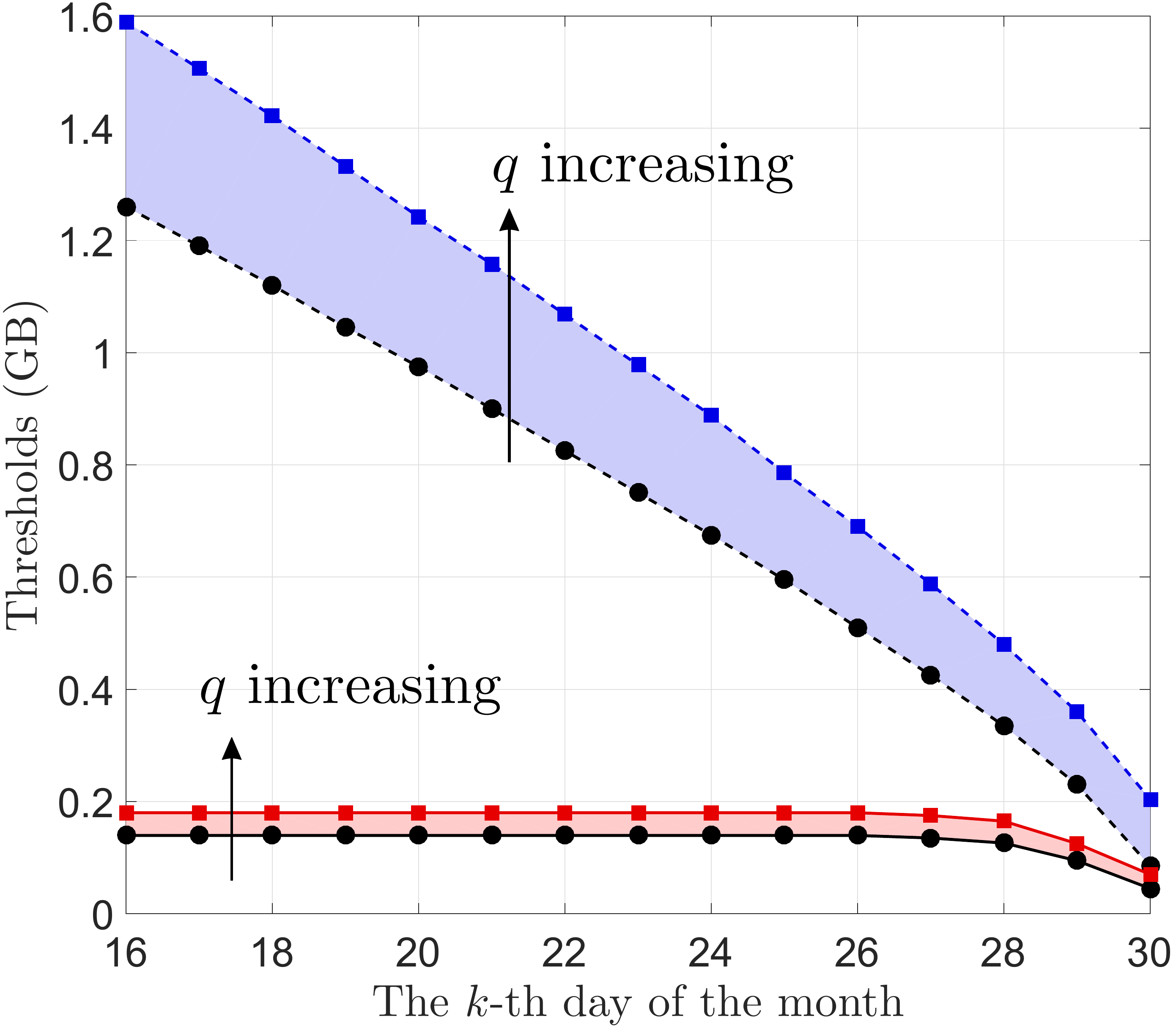}}
	\caption{Rollover-involved trading thresholds with $\discount=0.98$.}
	\label{fig: Threshold range}	
\end{figure}

We observe from Fig. \ref{fig: Threshold_Pure} that a larger time discount leads to the increase of both threshold values.
That is, the user tends to \textit{sell less} and \textit{buy more} data as $\discount$ increases.
This is because that a larger discount $\discount$ corresponds to a better joint consideration for the current and the future, which is twofold: 
\begin{itemize} 
	\item The user is willing to be  \textit{more patient} to sell data for immediate income.
	\item The user is \textit{less sensitive} to incur immediate cost from buying data.
\end{itemize}

\begin{figure*}
	\centering 
	\setlength{\abovecaptionskip}{0pt}
	\setlength{\belowcaptionskip}{0pt}
	\subfigure[Average daily trading prices.]{\label{fig: Test_DailyPrice}\includegraphics[height=0.22\linewidth]{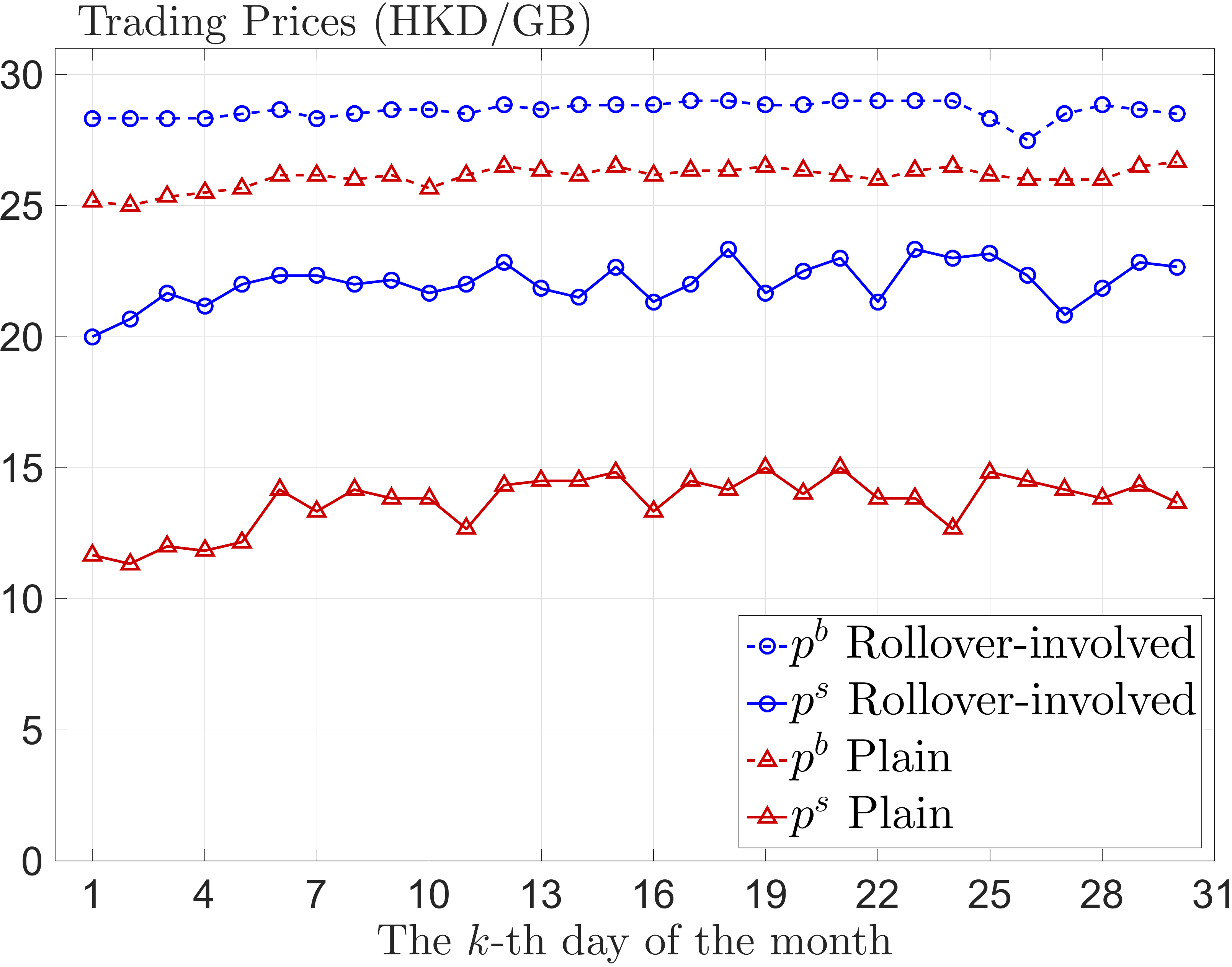}} \qquad
	\subfigure[Average user payoff.]{\label{fig: Test_Payoff}\includegraphics[height=0.22\linewidth]{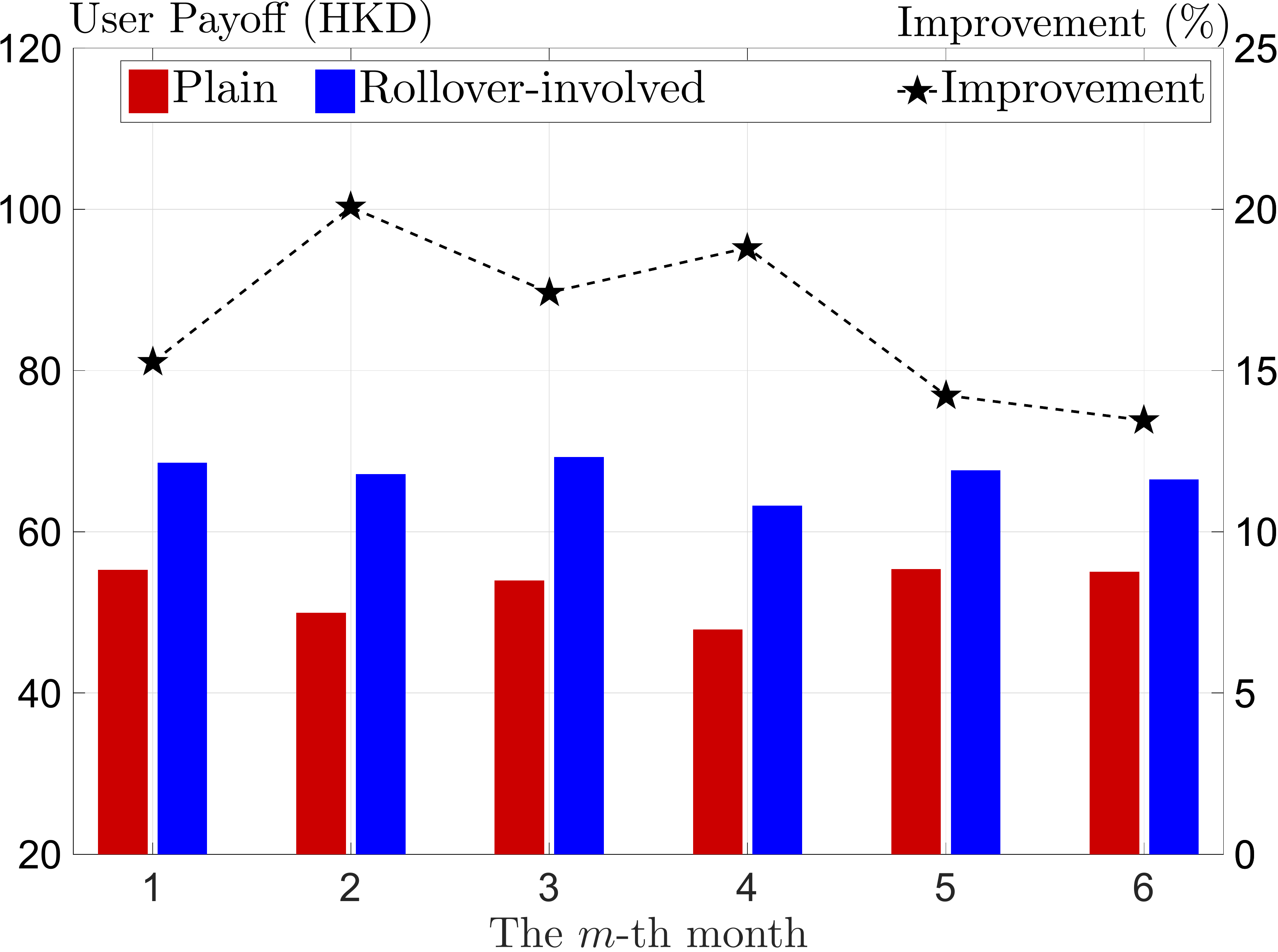}}  \qquad
	\subfigure[MNO's revenue from one user.]{\label{fig: Test_Revenue}\includegraphics[height=0.22\linewidth]{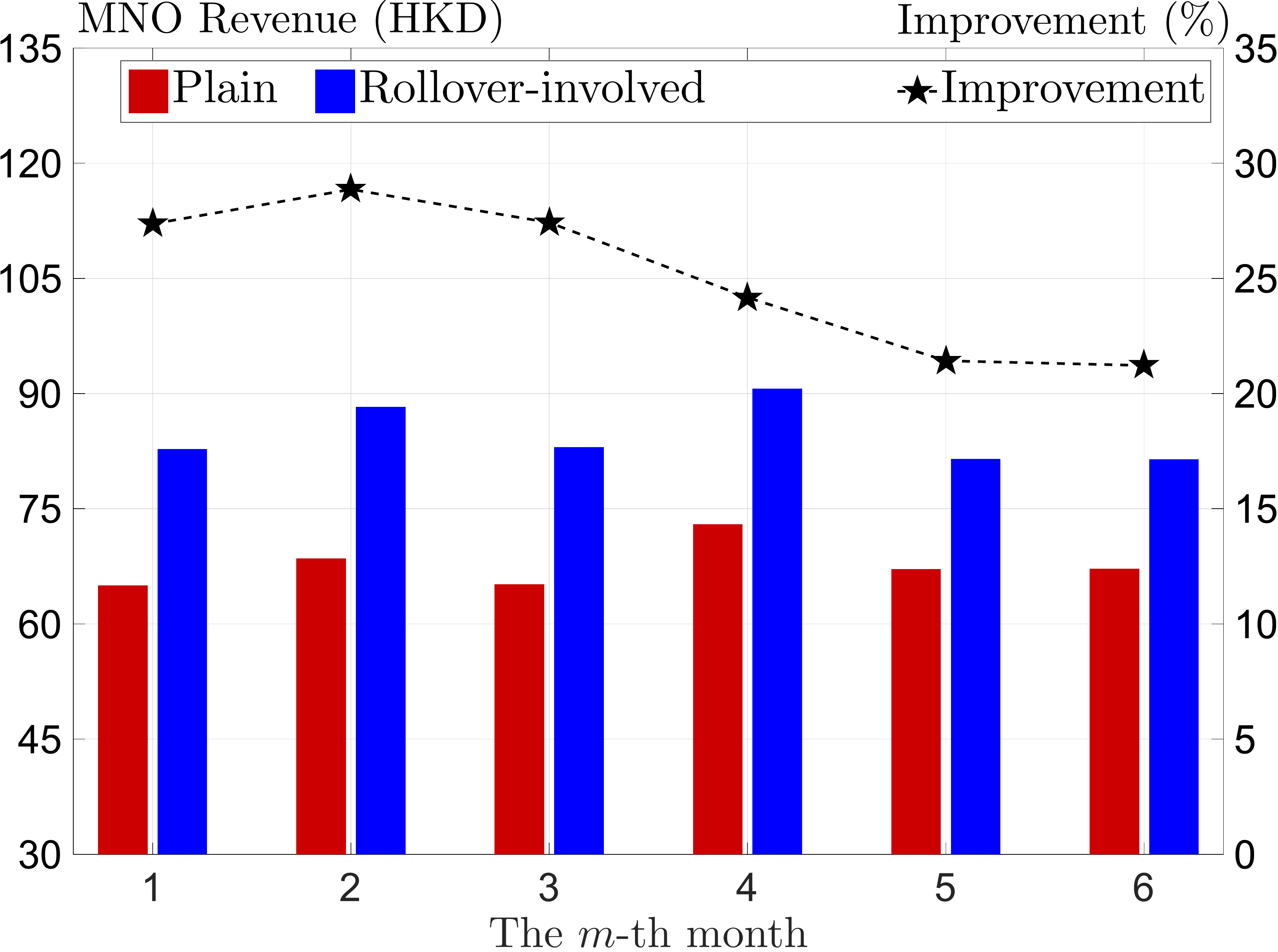}} 
	\caption{Performance evaluation.\vspace{-5pt}}
	\label{fig: Evaluation}
\end{figure*}
\subsubsection{Rollover-involved Trading}
Now we look at the rollover-involved trading scenario.
Recall that the buy-up-to threshold $L^\text{Roll}_{k,m}$ and the sell-down-to threshold $U^\text{Roll}_{k,m}$ are related to the long-term data volume $q$.
%Moreover, the rollover-involved trading policies are different even for the same day of different months.
%In the following, we investigate the impact of long-term data volume $q$ on the thresholds.
%Due to space limit, here we will take User 1 as example.
Fig. \ref{fig: Threshold range} shows how the long-term data volume $q$ affects the two thresholds with $\discount=0.98$.
Similarly, the horizontal axis represents the $k$-th day in a month.
The two black circle curves represent the two thresholds with zero long-term data, i.e., $L^\text{Roll}_{k,m}(p^b,0)$ and $U^\text{Roll}_{k,m}(p^s,0)$.
The two red square curves are the thresholds with the maximal long-term data, i.e., $L^\text{Roll}_{k,m}(p^s,\dcap)$ and $U^\text{Roll}_{k,m}(p^b,\dcap)$.
Therefore, the buy-up-to threshold $L^\text{Roll}_{k,m}(p^b,q)$ will appear in the red region and sell-down-to threshold $U^\text{Roll}_{k,m}(p^b,q)$ will appear in the blue region.
On average, the long-term data volume $q$ leads to $20\%$ and $10\%$ increase of the buy-up-to and sell-down-to threshold values, respectively.

%
%Fig. \ref{fig: Threshold_Roll Q} compares the selling and buying curves in the same day of different months.
%The two sub-figures correspond to different monthly data caps.
%In each sub-figure, the horizontal and vertical axises represent the long-term and short-term data volume, respectively.
%We plot three buying curves and three selling curves in each sub-figure.
%By comparing Fig. \ref{fig: Threshold_Roll_Q1_User1} and Fig. \ref{fig: Threshold_Roll_Q2_User1}, we note that a larger data cap draws closer the buying (and selling) curves in different months.
%This is because that users do not need to consider the future months too much if the monthly data cap $\dcap$ is large enough.
%Hence the policies of different months (except the last) are more homogeneous.

%\begin{figure} 
%	\centering 
%	\setlength{\abovecaptionskip}{0pt}
%	\setlength{\belowcaptionskip}{0pt}
%	\subfigure[Average trading prices]{\label{fig: Test_DailyPrice}\includegraphics[height=0.43\linewidth]{Test_DailyPrice.png}} \ \ 
%	\subfigure[MNO's revenue from one user.]{\label{fig: Test_Revenue}\includegraphics[height=0.43\linewidth]{Test_Revenue.png}} 
%	\caption{Performance evaluation.}
%	\label{fig: Evaluation}
%\end{figure}

\subsection{Performance Evaluation}\label{Subsection: Performance Evaluation}
To quantitatively evaluate the effect of rollover mechanism, we consider $\dcap=1$GB data cap with $\pcap=100$HKD monthly subscription fee.
The per-unit fee is $\adfee=30$HKD/GB.
Based on the estimated daily data consumption distribution, we randomly generate six-month daily data usage for 500 mobile users.
We consider two cases with and without the rollover mechanism.
For each case, we first determine whether the user will subscribe to this data plan based on his value function, then we simulate the MNO's pricing, users' trading and data consumption process for six months.

Fig. \ref{fig: Test_DailyPrice} plots the average daily trading prices in a month.
The two red curves (marked by triangles) represent the trading prices without rollover mechanism, while the two blue curves (marked by circles) represent the case with rollover mechanism.
We note that the rollover mechanism drives both the buying and selling prices higher.
This is because that the time-flexibility enables users to buy more but sell less data, leading to a seller's market with higher trading prices.
Fig. \ref{fig: Test_Payoff} and Fig. \ref{fig: Test_Revenue} plot users' average monthly payoff and the MNO's average monthly revenue, respectively.
We observe from the black curves (marked by stars) that the rollover mechanism can increase users' monthly payoff by 17\% on average  and the MNO's monthly revenue by 25\% on average, compared with the case without rollover mechanism.
These improvements are quite substantial.

\section{Conclusion} \label{Section: conclusion}
In this paper, we studied the economic viability of the data trading market with the rollover mechanism and investigated the interrelationship between the time-flexibility and user-flexibility.
We found that the time-flexible rollover mechanism benefits the user-flexible data trading market, in the sense that it can substantially increases both users' expected payoff and the MNO's expected revenue.

%Since this is the first work studying both time-flexibility and user-flexibility, there are still some open problems not comprehensively studied.
%In the future, we will further consider the following aspects.
%\begin{itemize} 
%	\item First, we will study the MNO's long-term revenue maximization problem.
%			In this case, the MNO may not always clear the market in each day, but will consider the long-term impact of the price choices by predicting users' future consumption patterns and trading decisions.  
%	\item Second, we will take into account users' on-line learning process on the trading prices.
%			A user's belief on the future trading prices may update according to the price history.
%			It is interesting and practically relevant to consider this process and investigate its effect on the MNO's revenue.
%	\item Third, we will consider the bounded cognitive capacity of users.
%			In practice, users may not explicitly trade data according to our optimal trading policy due to the limited knowledge or computational power, but subconsciously adjust the data inventory based on the prices and the remaining days.
%			It would be interesting to consider how such behaviors affect the aggregate demand and supply. 
%\end{itemize} 

\bibliographystyle{IEEEtran}
\bibliography{ref}

\begin{thebibliography}{10}
\providecommand{\url}[1]{#1}
\csname url@samestyle\endcsname
\providecommand{\newblock}{\relax}
\providecommand{\bibinfo}[2]{#2}
\providecommand{\BIBentrySTDinterwordspacing}{\spaceskip=0pt\relax}
\providecommand{\BIBentryALTinterwordstretchfactor}{4}
\providecommand{\BIBentryALTinterwordspacing}{\spaceskip=\fontdimen2\font plus
\BIBentryALTinterwordstretchfactor\fontdimen3\font minus
  \fontdimen4\font\relax}
\providecommand{\BIBforeignlanguage}[2]{{%
\expandafter\ifx\csname l@#1\endcsname\relax
\typeout{** WARNING: IEEEtran.bst: No hyphenation pattern has been}%
\typeout{** loaded for the language `#1'. Using the pattern for}%
\typeout{** the default language instead.}%
\else
\language=\csname l@#1\endcsname
\fi
#2}}
\providecommand{\BIBdecl}{\relax}
\BIBdecl

\bibitem{sen2013survey}
S.~Sen, C.~Joe-Wong, S.~Ha, and M.~Chiang, ``A survey of smart data pricing:
  Past proposals, current plans, and future trends,'' \emph{ACM Computing
  Surveys (CSUR)}, vol.~46, no.~2, p.~15, 2013.

\bibitem{ATTrollover}
\emph{AT\&T Rollover Data Plan}, \url{https://www.att.com}.

\bibitem{CUHKrollover}
\emph{China Unicom HK Rollover Data Plan}, \url{https://www.cuniq.com/hk}.

\bibitem{CMrollover}
\emph{China Mobile Rollover Data Plan}, \url{http://www.10086.cn}.

\bibitem{Zhiyuan2018TMC}
\BIBentryALTinterwordspacing
Z.~{Wang}, L.~{Gao}, and J.~{Huang}, ``Exploring time flexibility in wireless
  data plans,'' \emph{IEEE Transactions on Mobile Computing}, 2018. [Online].
  Available: \url{https://arxiv.org/abs/1808.10569}
\BIBentrySTDinterwordspacing

\bibitem{CMHK2cm}
\emph{China Mobile Hong Kong, 2cm}, \url{http://www.hk.chinamobile.com}.

\bibitem{zheng2015secondary}
L.~Zheng, C.~Joe-Wong, C.~W. Tan, S.~Ha, and M.~Chiang, ``Secondary markets for
  mobile data: Feasibility and benefits of traded data plans,'' in \emph{Proc.
  of IEEE INFOCOM}, 2015.

\bibitem{ha2012tube}
S.~Ha, S.~Sen, C.~Joe-Wong, Y.~Im, and M.~Chiang, ``Tube: time-dependent
  pricing for mobile data,'' in \emph{Proc. of ACM SIGCOMM}, 2012.

\bibitem{ma2016time}
Q.~Ma, Y.~Liu, and J.~Huang, ``Time and location aware mobile data pricing,''
  \emph{IEEE Transactions on Mobile Computing}, vol.~15, no.~10, pp.
  2599--2613, 2016.

\bibitem{ma2016usage}
R.~T. Ma, ``Usage-based pricing and competition in congestible network service
  markets,'' \emph{IEEE/ACM Transactions on Networking}, vol.~24, no.~5, pp.
  3084--3097, 2016.

\bibitem{andrews2014understanding}
M.~Andrews, G.~Bruns, M.~Do{\u{g}}ru, and H.~Lee, ``Understanding quota
  dynamics in wireless networks,'' \emph{ACM Transactions on Internet
  Technology (TOIT)}, vol.~14, no. 2-3, p.~14, 2014.

\bibitem{Zheng2018dynamics}
L.~Zheng, C.~Joe-Wong, M.~Tan, Andrews, and M.~Chiang, ``Optimizing data plans
  usage dynamics in mobile data networks,'' in \emph{Proc. of IEEE INFOCOM},
  2018.

\bibitem{zheng2015customized}
L.~Zheng, C.~Joe-Wong, C.~W. Tan, S.~Ha, and M.~Chiang, ``Customized data plans
  for mobile users: Feasibility and benefits of data trading,'' \emph{IEEE
  Journal on Selected Areas in Communications}, vol.~35, no.~4, pp. 949--963,
  2017.

\bibitem{yu2017mobile}
J.~Yu, M.~H. Cheung, J.~Huang, and H.~V. Poor, ``Mobile data trading:
  Behavioral economics analysis and algorithm design,'' \emph{IEEE Journal on
  Selected Areas in Communications}, vol.~35, no.~4, pp. 994--1005, 2017.

\bibitem{andrews2016understanding}
M.~Andrews, ``Understanding the effects of quota trading on mobile usage
  dynamics,'' in \emph{Proc. of IEEE WiOpt}, 2016.

\bibitem{zheng2016understanding}
L.~Zheng and C.~Joe-Wong, ``Understanding rollover data,'' in \emph{Smart Data
  Pricing Workshop}, 2016.

\bibitem{wei2018novel}
\BIBentryALTinterwordspacing
Y.~Wei, J.~Yu, T.~M. Lok, and L.~Gao, ``A novel mobile data contract design
  with time flexibility,'' \emph{IEEE Transactions on Mobile Computing}, 2018.
  [Online]. Available: \url{https://arxiv.org/abs/1806.07308}
\BIBentrySTDinterwordspacing

\bibitem{Zhiyuan2017rollover}
Z.~Wang, L.~Gao, and J.~Huang, ``Pricing optimization of rollover data plan,''
  in \emph{Proc. of IEEE WiOpt}, 2017.

\bibitem{Zhiyuan2018duopoly}
Z.~{Wang}, L.~Gao, and J.~Huang, ``Pricing competition of rollover data plan,''
  in \emph{Proc. of IEEE WiOpt}, 2018.

\bibitem{Zhiyuan2018MobiHoc}
Z.~Wang, L.~{Gao}, and J.~Huang, ``Multi-dimensional contract design for mobile
  data plan with time flexibility,'' in \emph{Proc. of ACM MobiHoc}, 2018.

\bibitem{villas2004consumer}
J.~M. Villas-Boas, ``Consumer learning, brand loyalty, and competition,''
  \emph{Marketing Science}, vol.~23, no.~1, pp. 134--145, 2004.

\bibitem{Technicalreport}
On-line technical report.
  \url{https://www.dropbox.com/s/ib3hhw1prbiodje/Infocom19.pdf?dl=0}.

\bibitem{InternetUsage}
\emph{Analysis and modeling of Internet usage},
  \url{https://www.kaggle.com/andrewfager/analysis-and-modeling-of-internet-usage}.

\end{thebibliography}

\end{document}